\documentclass[a4paper,11pt]{article}
\usepackage{jheppub}
\usepackage{style}

\preprint{FERMILAB-PUB-25-0892-SQMS-T}

\title{Infrared Freeze-In of Magnetic Dipole Dark Matter}

\author[a,b]{Asher Berlin\,\orcidlink{0000-0002-1156-1482},}
\author[a,c]{Jae Hyeok Chang\,\orcidlink{0000-0003-2769-3858}}
\author[a,d]{and Tanner Trickle\,\orcidlink{0000-0003-1371-4988}}

\affiliation[a]{Theoretical Physics Division, Fermi National Accelerator Laboratory, Batavia, IL 60510, USA}
\affiliation[b]{Superconducting Quantum Materials and Systems Center (SQMS), Fermi National Accelerator Laboratory, Batavia, IL 60510, USA}
\affiliation[c]{Department of Physics, University of Illinois Chicago, Chicago, IL 60607, US}
\affiliation[d]{Department of Physics, Grainger College of Engineering, University of Illinois Urbana-Champaign, Urbana, IL 61801, USA}

\emailAdd{aberlin@fnal.gov}
\emailAdd{jhchang@fnal.gov}
\emailAdd{ttrickle@illinois.edu}

\abstract{We propose a novel mechanism for the cosmological production of keV -- GeV mass dark matter that interacts with the Standard Model through a small effective magnetic dipole moment. Such an interaction can be radiatively generated if dark matter couples to heavier charged particles. Previous studies have focused on the case where these charged states are much heavier than the reheat temperature, such that freeze-in production of dark matter is sensitive to the ultraviolet details of reheating. Here, we instead consider the possibility that these heavy states have masses comparable to the dark matter mass and are charged under a new kinetically-mixed $U(1)'$. As a result, dark matter production is dominated by the infrared freeze-in of the heavy charged states that subsequently thermalize the rest of the dark sector to a temperature much below that of the visible bath. We delineate regions of parameter space consistent with cosmological and astrophysical constraints and identify benchmark scenarios that can guide the next generation of direct detection experiments searching for spin-dependent scattering of sub-GeV dark matter.}

\begin{document}
\maketitle
\newpage

\section{Introduction}
\label{sec:introduction}
The absence of evidence for new particles with electroweak scale masses from direct detection, indirect detection, and collider experiments has sparked new ideas regarding what the particle nature of dark matter (DM) might be. DM candidates with keV -- GeV masses, often referred to as ``light" DM, are particularly compelling candidates~\cite{Knapen:2017xzo,Lin:2019uvt,Zurek:2024qfm}. Indeed, many of the theoretically well-motivated production mechanisms for light DM predict detectably large couplings to the Standard Model (SM). These include examples where MeV -- GeV mass DM was in thermal equilibrium with the SM in the early universe, for instance where the relic abundance is established from the freeze-out of annihilations to visible final states or processes involving solely dark sector (DS) species~\cite{Battaglieri:2017aum}. Lighter sub-MeV mass DM candidates could also have been produced from thermal freeze-out if equilibrium was achieved late, after neutrino-photon decoupling~\cite{Berlin:2017ftj,Berlin:2018ztp}. 

Additionally, light DM could have been produced by out-of-equilibrium thermal processes.  In this case, the DM-SM coupling is too weak to thermalize the two sectors, yet is strong enough to ``freeze-in" a sizable DM abundance~\cite{Hall:2009bx,Chu:2011be}. If the initial DM abundance is negligible, these out-of-equilibrium processes can generate a DM density consistent with observations. A minimal example allowing for such a cosmology involves a single DM particle with a vector-portal-coupling to electromagnetically-charged SM species. This simple model has become one of the standard benchmarks and has long been at the edge of detectability for electron-based direct detection experiments searching for the spin-independent scattering of sub-GeV DM, such as CDEX~\cite{CDEX:2019exx,CDEX:2022kcd}, DAMIC~\cite{DAMIC-M:2023gxo,DAMIC-M:2025luv}, EDELWEISS~\cite{EDELWEISS:2020fxc}, SENSEI~\cite{SENSEI:2023zdf},  SuperCDMS~\cite{SuperCDMS:2019jxx,SuperCDMS:2020ymb}, as well as the proposed Oscura experiment~\cite{Oscura:2022vmi}. 

Notably, the sensitivity required to directly probe this model has recently been achieved by the DAMIC collaboration~\cite{DAMIC-M:2025luv}, which has ruled out this freeze-in benchmark in the $\sim$ 1 MeV -- 100 MeV mass range. In light of this recent progress, there is strong motivation to further explore the landscape of light DM beyond the standard freeze-in target, in order to determine if remaining models allow for a wider variety of low-energy phenomenology, such as interactions that are primarily spin-dependent~\cite{Berlin:2023ubt,Gori:2025jzu}. Variations to the low-energy phenomenology are also relevant to the next generation of direct detection experiments utilizing collective excitations, such as phonons~\cite{Knapen:2017ekk,Griffin:2018bjn,Trickle:2019nya,Kurinsky:2019pgb,Cox:2019cod,Coskuner:2021qxo,Trickle:2020oki,Griffin:2019mvc,Knapen:2021bwg,Griffin:2020lgd,Taufertshofer:2023rgq,Raya-Moreno:2023fiw} and magnons~\cite{Trickle:2019ovy,Trickle:2020oki,Esposito:2022bnu,Marocco:2025eqw,Berlin:2025uka}, since the response of these detectors can depend dramatically on the spin-dependence of the interaction~\cite{Trickle:2020oki}. 

In this work, we explore one such variation where fermionic DM  couples to the SM through a dark magnetic dipole moment (MDM). Previous work has shown that DM may be produced from freeze-in directly from the magnetic dipole interaction~\cite{Chang:2019xva}. However, due to the higher-dimensional nature of the interaction, freeze-in production is dominated at the highest temperatures in this case~\cite{Elahi:2014fsa}, implying that the only detectable parameter space involves scenarios where the reheat temperature of the universe is extremely small. In contrast, we focus on a production mechanism that does not have this feature and is instead UV-\textit{insensitive} to the same degree as the usual IR freeze-in scenario. 

The central idea of our work is that the same particles responsible for radiatively generating the MDM can acquire an initial abundance from freeze-in that subsequently thermalizes the rest of the DS. A minimal concrete model realizing these dynamics involves multiple DS states: the neutral DM candidate $\chi$, and heavier dark charged particles (DCPs) $\psi$ and $\phi$ that are directly charged under a kinetically-mixed dark photon $A'$. The SM populates an initially small abundance of DCPs through freeze-in processes involving the kinetic mixing. Due to sizable couplings between DS species, this is followed by number-changing reactions that fully thermalize the rest of the DS to a temperature much smaller than that of the SM bath. At later times, $\psi$ and $\phi$ deplete their thermal abundance through, e.g., annihilations to $\chi$ and $A'$. Finally, $\chi$ ultimately decouples from the rest of the DS as either a hot or cold relic, depending on the DS temperature and size of the $\chi$-$A'$ interaction. Although the cosmological abundance of DCPs is negligible at late times, they play the important role of radiatively generating $\chi$'s magnetic dipole, providing the dominant interaction with the SM for direct detection experiments. We find that for sizable DS couplings, and $\mathcal{O}(1)$ DCP-DM mass ratios, there is astrophysically and cosmologically viable parameter space near the edge of detectability for future experiments searching for MeV mass DM (Fig.~\ref{fig:multi-panel}). 

This paper is organized as follows. In Sec.~\ref{sec:model}, we present an example model whose low-energy phenomenology is determined by a dark MDM. In Sec.~\ref{sec:cosmology}, we derive the cosmological history of our example model, which involves a detailed understanding of DCP freeze-in (Sec.~\ref{subsec:dark_sector_FI}), DS thermalization (Sec.~\ref{subsec:thermalization}), the DS energy density evolution (Sec.~\ref{subsec:DS_temperature_evolution}), and determining how the DS particles decouple from the DS thermal bath (Sec.~\ref{subsec:ds_particle_decouple}). In Sec.~\ref{sec:astrophysics}, we derive limits from the consideration of cosmological and astrophysical processes, such as changes to the effective number of neutrino species, structure formation, and stellar energy loss. In Sec.~\ref{sec:target_parameter_space}, we highlight viable parameter space motivated by our cosmological scenario and discuss MDM signals in current and future direct detection experiments. Finally, we conclude in Sec.~\ref{sec:conclusions}.

\section{An Example Model}
\label{sec:model}

Here, we discuss an example model whose Lagrangian can be split into three components: $\mathcal{L} = \mathcal{L}_\text{SM} + \mathcal{L}_\text{DS} + \mathcal{L}_\text{DS-SM}$, where $\mathcal{L}_\text{SM}$ is the SM Lagrangian, $\mathcal{L}_\text{DS}$ is the DS Lagrangian, and $\mathcal{L}_\text{DS-SM}$ is the Lagrangian which couples the DS to the SM. We begin by introducing a dark $U(1)'$ gauge symmetry in the DS with an associated dark photon, $A'$, that couples to the SM via kinetic mixing,
\begin{align}
    \mathcal{L}_{\text{DS}-\text{SM}} & = - \frac{\varepsilon}{2} \, F^{\mu \nu} F'_{\mu \nu} \, ,
    \label{eq:L_DS-SM}
\end{align}
where $F_{\mu \nu}^{(\prime)} = \partial_\mu A^{(\prime)}_\nu - \partial_\nu A_\mu^{(\prime)}$ is the SM (DS) electromagnetic field strength tensor, and $\varepsilon$ is the kinetic mixing parameter.\footnote{At higher energies, the relevant interaction in Eq.~\eqref{eq:L_DS-SM} is between $F'_{\mu \nu}$ and the hypercharge field strength tensor. Full expressions can be found in App.~\ref{app:integrated_density_collision_terms}.} To leading order in $\varepsilon \ll 1$, the kinetic terms can be brought into canonical form by the field redefinition $A_\mu \rightarrow A_\mu - \varepsilon A'_\mu$. This yields the interaction term $\mathcal{L} \supset -\varepsilon e \, q_\Psi \, A'_\mu \bar{\Psi} \gamma^\mu \Psi$ (with $e = -|e|$), which couples the $A'$ to all electromagnetically-charged SM states, $\Psi$, with electromagnetic charge $q_\Psi$ (e.g., $q_e = -1$ for the electron).

The canonical freeze-in cosmology takes the DM, $\chi$, to be directly charged under this $U(1)'$, such that the $A'$ directly mediates DM-SM interactions. In this work, we instead assume that $\chi$ is a Dirac fermion that is \emph{uncharged} under the $U(1)'$. In this case, the $\chi$-$A'$ interaction is radiatively generated by heavier DS particles charged under the $U(1)'$. We take these to be a complex scalar, $\phi$, and a Dirac fermion, $\psi$, referred to throughout as ``dark charged particles" (DCPs). The interactions within the DS are described by
\begin{align}
    \mathcal{L}_{\text{DS}} = &- \frac{1}{4} F'_{\mu \nu} {F'}^{\mu \nu} + \left| D_\mu' \phi \right|^2- m_\phi^2 |\phi|^2 + \bar{\psi} (i \slashed{D}' - m_\psi) \psi + \bar{\chi} (i \slashed{\partial} - m_\chi) \chi 
    \nonumber \\
    &+ y' \left( \phi \bar{\psi} \chi + \phi^* \bar{\chi} \psi \right) \, ,
    \label{eq:L_DS}
\end{align}
where $D'_\mu = \partial_\mu + i g' q' A'_\mu$, $m_\psi$, $m_\phi$, and $m_\chi$ are the $\psi \,, \phi$, and $\chi$ masses, respectively, $g' = |g'|$ is the DS gauge coupling, $y'$ is the DS Yukawa that couples DM to the DCPs, and the DCPs are assumed to have charge $q' = +1$ under the $U(1)'$. After performing the photon field redefinition to remove the kinetic mixing in Eq.~\eqref{eq:L_DS-SM}, the phenomenologically relevant interactions are
\begin{align}
    \mathcal{L} \supset -A'_\mu \left( g' J_\phi^{\prime\mu} + g' J_\psi^{\prime\mu} + \varepsilon e \, J_\text{SM}^\mu \right) + g'^2 |\phi|^2 A'_\mu A^{\prime \mu} + y' \left( \phi \bar{\psi} \chi + \phi^* \bar{\chi} \psi \right)
    \, ,
    \label{eq:int_L}
\end{align}
where $-iJ_\phi^{\prime\mu} = \phi^* (\partial^\mu \phi) - \phi \, (\partial^\mu \phi^*)$ and $J_\psi^{\prime\mu} = \bar{\psi} \gamma^\mu \psi$ are the $U(1)'$ charge currents, and $J_\text{SM}^\mu = q_e \, \bar{e} \gamma^\mu e + q_\mu \, \bar{\mu} \gamma^\mu \mu + \cdots$ is the total electromagnetic SM current.

\begin{figure}
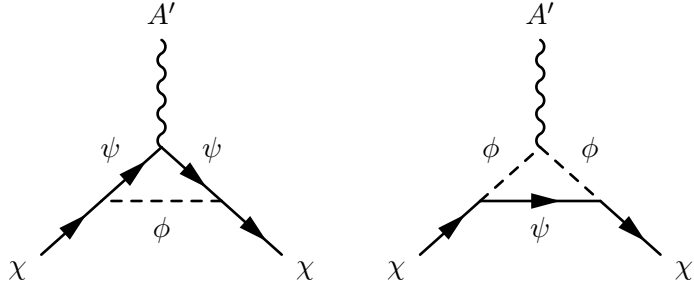

    \centering
    \include{feynman_diagrams/chi_mdm_diagram}
    \vspace{-3em}
    \caption{Feynman diagrams contributing to the dark matter, $\chi$, dark magnetic dipole moment, $\mu'_\chi$ (Eq.~\eqref{eq:L_EFT}).}
    \label{fig:chi_mdm_feynman_diagrams}
\end{figure}

Observables in direct detection experiments are determined by how $\chi$ couples to the $A'$. The leading order interaction is due to a loop of DCPs, shown in Fig.~\ref{fig:chi_mdm_feynman_diagrams}, which generates a dark MDM for the DM. At low energies, the effective field theory 
of DS-SM interactions is then
\begin{align}
    \mathcal{L}_\text{EFT} \supset -\varepsilon e \, A'_\mu J_\text{SM}^\mu - \frac{\mu_\chi'}{2} \, \bar{\chi} \sigma^{\mu \nu} \chi \, F'_{\mu \nu} \, ,
    \label{eq:L_EFT}
\end{align}
where $\sigma^{\mu \nu} = i [\gamma^\mu, \gamma^\nu] / 2$ and $\mu'_\chi$ is the DM's dark MDM, determined by evaluating the diagrams in Fig.~\ref{fig:chi_mdm_feynman_diagrams}.\footnote{For approximately massless dark photons, we can alternatively redefine the dark photon field as $A'_\mu \rightarrow A'_\mu - \varepsilon A_\mu$. With this choice, $\psi$ and $\phi$ become millicharged under the SM photon with charge $\varepsilon g'$, and $\chi$ gains an effective MDM $\mu_\chi = \varepsilon \mu_\chi'$ under the visible sector.\label{foot}}

 We take the DM mass to be $m_\chi \lesssim m_\psi / 3$.\footnote{We avoid $m_\chi / M$ mass ratios which are smaller than the loop-generated value, $m_\chi / M \sim y'^2 / 16 \pi^2$~\cite{Graham:2012su}.} We also assume a mass hierarchy of $m_\phi > m_\psi + m_\chi$ to allow for the decay $\phi \rightarrow \psi + \chi$. In order to simplify the notation further, in all expressions unrelated to $\phi \rightarrow \psi + \chi$ decays, we assume $m_\phi \simeq m_\psi \equiv M$, where $M$ is the DCP mass scale. In this limit, the DM  MDM generated by the diagrams in Fig.~\ref{fig:chi_mdm_feynman_diagrams} is~\cite{Cheung:2009fc,Graham:2012su}
\begin{align}
    \mu'_\chi = \frac{g' y'^2}{32 \pi^2 M} \, .
    \label{eq:dark_magnetic_dipole_moment}
\end{align}
We will later find it useful to introduce the $\kappa$ parameter in analogy with Refs.~\cite{Chu:2011be, Bhattiprolu:2023akk,Bhattiprolu:2024dmh},
\begin{align}
    \kappa \equiv \frac{\varepsilon g'}{e} \, .
    \label{eq:kappa}
\end{align}

\section{Cosmological History}
\label{sec:cosmology}

\begin{figure}[t!]
    \centering
    \includegraphics[width=\linewidth]{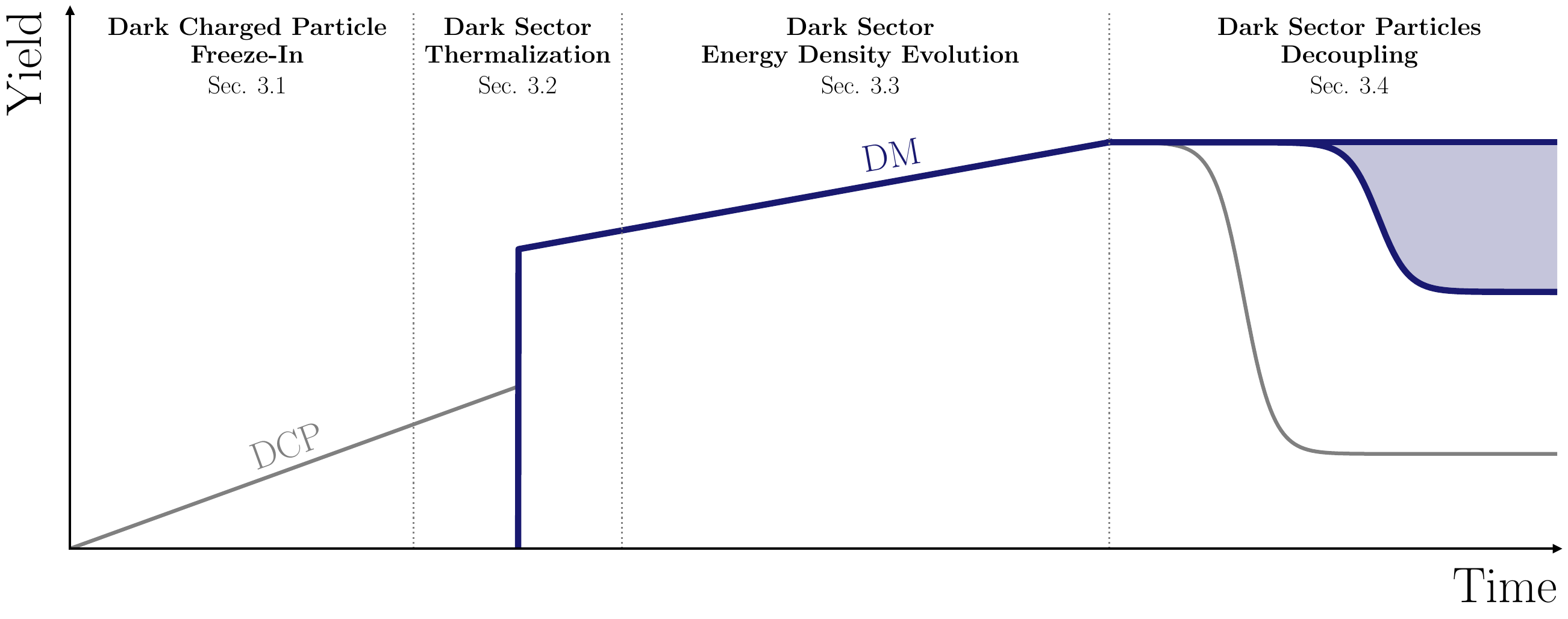}
    \caption{Schematic evolution of the dark matter (DM) and dark charged particle (DCP) yields throughout the cosmological epochs discussed in Sec.~\ref{sec:cosmology}. The solid gray line tracks the DCP yield, the blue line tracks the DM yield, and the shaded blue region represents the range of DM relic abundances depending on how strongly the DM couples to the rest of the DS. At early times, the DCPs freeze-in from the SM plasma (Sec.~\ref{subsec:dark_sector_FI}) until internal DS reactions are strong enough to thermalize the rest of the DS (Sec.~\ref{subsec:thermalization}). After DS thermalization, each species follows an equilibrium distribution set by the DS temperature, which is continually increasing relative to the SM temperature due to continued energy injection (Sec.~\ref{subsec:DS_temperature_evolution}). Lastly, each of these particle species decouples from the DS. The order in which these events occur can parametrically alter the final DS abundances (Sec.~\ref{subsec:ds_particle_decouple}).}
    \label{fig:abundance_cartoon}
\end{figure}

In this section, we begin with a brief overview of various DS cosmological histories, illustrated in Fig.~\ref{fig:abundance_cartoon}, in order to provide context for the following subsections. We assume that the yields of all particles in the DS are negligible at early times, such that the initial DS abundance is composed of DCPs which are frozen-in from interactions with SM charged particles (Sec.~\ref{subsec:dark_sector_FI}). This is analogous to the minimal freeze-in scenario~\cite{Hall:2009bx}; however, since here the DCPs are \textit{not} the DM, a different mechanism is needed to generate the DM abundance~\cite{Chu:2011be,Fernandez:2021iti,Bhattiprolu:2023akk,Bhattiprolu:2024dmh}. This automatically occurs if the DS couplings are sufficiently large. In this case, once the frozen-in DCP abundance is sufficiently large, DM production arises from thermalization within the DS, which creates a DS bath at temperature $T'$ which is much less than that of the the SM, $T$ (Sec.~\ref{subsec:thermalization}). After this occurs, all of the DS particles thus follow an equilibrium distribution at temperature $T'$. Later, each of the DS particles eventually decouple from the DS bath and their comoving density is frozen out. 

Due to the mass-hierarchy and interaction structure introduced above, the DCPs will freeze-out within the DS to a negligible abundance at late times.  On the other hand, $\chi$ can decouple as a hot or cold relic with an abundance in agreement with the observed DM energy density, depending on the sizes of various DS couplings as well as the DCP-to-DM mass ratio (blue shaded region in Fig.~\ref{fig:abundance_cartoon}). 

The order in which these events happen, relative to when energy injection from the SM to the DS decouples, parametrically changes the final yield of various DS species. We therefore split our discussion of DS decoupling in Sec.~\ref{subsec:ds_particle_decouple} into two different subsections (Secs.~\ref{subsubsec:scenario_I} and \ref{subsubsec:scenario_II}). 

\subsection{Dark Charged Particle Freeze-In}
\label{subsec:dark_sector_FI}

The DS cosmological evolution begins when the DCPs acquire an initial abundance from SM-to-DS freeze-in. While the discussion of freeze-in is relatively standard (see, e.g., Refs.~\cite{Hall:2009bx,Chang:2019xva,Dvorkin:2019zdi}), we present it here in order to be self-contained and introduce notation that will be useful later.

At early times, the DCP yields are much smaller than their equilibrium value at the SM temperature $T$, i.e., $Y_X \ll Y_X^\text{eq}$, where $X$ indexes the DCPs ($X \in \{ \psi, \phi\}$), $Y_X = n_X / s$, $Y^\text{eq}_X = n_X^\text{eq} / s$, $n_X$ and $n_X^\text{eq}$ are the number density and equilibrium number density (at temperature $T$) of $X$, and $s$ is the entropy density. The rate for SM-to-DCP freeze-in processes, $\Gamma_{\text{SM} \to X}$, is too weak to establish equilibrium between the two sectors, corresponding to $\Gamma_{\text{SM} \to X} \ll H$, where $H \sim T^2 / M_\text{Pl}$ is the Hubble parameter and $M_\text{Pl} \simeq 1.22 \times 10^{19} \, \text{GeV}$ is the Planck mass. Hence, at early times, before DS densities have grown to the point of enabling number-changing reactions and $T \gg m_e, M$, the frozen-in DCP yield takes the form $Y_X \sim \Gamma_{\text{SM} \to X} / H \propto e^4 \kappa^2 \, M_\text{Pl} / T$. 

This can be seen explicitly from the Boltzmann equation governing the initial freeze-in of $Y_X$,
\begin{align}
    \frac{dY_X}{dT} & = -\frac{c}{s H T} \sum_{i = \gamma^*, e, \mu, \ldots} R_{i \rightarrow X}(T)\, ,
    \label{eq:Y_boltz_FI}
\end{align}
where
\begin{equation}
c \equiv 1 + \frac{1}{3} \, \frac{d \ln (s / T^3)}{d \ln T} 
\, ,
\end{equation}
$i$ indexes the plasmon and SM charged particle contributions to DCP freeze-in, and $R_{i \rightarrow X}$ are the integrated collision terms for various processes. As an example, the contribution to the yield of $\psi$ from electron-positron annihilations is
\begin{align}
    R_{e \rightarrow \psi}(T) = \frac{T}{16 \, (2 \pi)^5} \int_{s_\text{min}}^\infty \dd s \, \sqrt{s}  \sqrt{1 - \frac{4 m_e^2}{s}} \sqrt{1 - \frac{4 M^2}{s}} \, \mathscr{M}^2_{ee \rightarrow \psi\psi} \, K_1(\sqrt{s} / T) 
    \, ,
    \label{eq:R_e_psi}
\end{align}
where $s_\text{min} = 4 \, \text{max}\{ m_e^2, M^2 \}$, and $\mathscr{M}^2$ is the spin-summed and angularly-averaged matrix element,
\begin{align}
    \mathscr{M}^2 \equiv \frac{1}{4 \pi} \sum_{\text{spins}} \int |\mathcal{M}|^2 \dd\Omega = \frac{1}{t_+ - t_-} \sum_\text{spins} \int_{t_-}^{t_+}  |\mathcal{M}|^2 \, \dd t \, ,
    \label{eq:summed_averaged_M}
\end{align}
where $t_\pm = - s/2 + m_e^2 + M^2 \pm (s/2) \sqrt{1 - 4 m_e^2 / s} \sqrt{1 - 4 M^2 / s}$. For electron-positron annihilations, $ee \rightarrow A^{\prime *} \to \psi\psi$, the matrix element (Eq.~\eqref{eq:summed_averaged_M}) relevant for Eq.~\eqref{eq:R_e_psi} is,
\begin{align}
    \mathscr{M}^2_{ee\rightarrow \psi\psi} = \frac{16}{3} e^4 \kappa^2 \left(1 + \frac{2 m_e^2}{s} \right) \left(1 + \frac{2 M^2}{s} \right) \, .
    \label{eq:M_scr_ee_psi_psi}
\end{align}
For $T \gg \text{max}\{ m_e, M \}$, $R_{e \rightarrow \psi} \propto T^4$, as expected from dimensional analysis. Therefore, in this limit, the Boltzmann equation in Eq.~\eqref{eq:Y_boltz_FI} becomes $d Y_X / dT \propto -M_\text{Pl} / T^2$, and the DCP yield evolves as~\cite{Dvorkin:2019zdi}
\begin{align}
    Y_X \sim 10^{-4} \, e^4 \kappa^2 \, \frac{M_\text{Pl}}{T} \, ,
    \label{eq:Y_FI_final}
\end{align}
as expected. In general, such freeze-in processes are negligible once $T \ll \max \{ m_e, M \}$ and rates for $e e \to X X$ become exponentially suppressed, either due to a small electron density or strong final-state kinematic suppression. 

A more detailed analysis would determine $Y_X$ by including all the contributions to the integrated collision term before numerically integrating Eq.~\eqref{eq:Y_boltz_FI}. However, as we show below, the DM relic abundance does not depend on the exact value of $Y_X$ in Eq.~\eqref{eq:Y_FI_final}, only that it grows beyond some threshold to thermalize the rest of the DS at low temperatures. The parametric estimate of $Y_X$ in Eq.~\eqref{eq:Y_FI_final} is therefore sufficient. 

\subsection{Dark Sector Thermalization}
\label{subsec:thermalization}

The DCP abundances continue to grow as in Eq.~\eqref{eq:Y_FI_final} until their interactions are strong enough to self-thermalize the rest of the DS. This is possible once various interactions within the DS set the chemical potential of each species to nearly zero, which can be satisfied with a variety of different reaction-sets, such as
\begin{align}
    \phi \leftrightarrow \psi\chi~~~,~~~\phi\phi\leftrightarrow A'A'~~~,~~~\psi\psi \leftrightarrow A'A'~~~,~~~\phi\phi \leftrightarrow \chi\chi \, .
\end{align}
In addition to this one, there are seven other sets of reactions made from combinations of $\phi \leftrightarrow \psi\chi$, and three of the five $\{ \phi\phi\leftrightarrow A'A',\,\,\psi\psi \leftrightarrow A'A',\,\,\phi\phi \leftrightarrow \chi\chi,\,\,\psi\psi\leftrightarrow\chi\chi,\,\,\phi \phi \leftrightarrow \psi\psi \}$, including at least one $A'$ and one $\chi$. For each of these eight reaction-sets $k \in \{1, \dots,8 \}$, there is a SM temperature $T_\text{th}^k$ at which thermalization will occur, corresponding to the time at which all processes in the set become faster than Hubble expansion. Thermalization thus occurs once the fastest of these combinations is allowed, i.e., $T_\text{th} = \max_k\{ T_\text{th}^k \}$. A precise determination of these thermalization temperatures requires knowledge of the out-of-equilibrium distribution functions of the DCPs. In this work, we instead use rough estimates and provide details of our approximations in App.~\ref{app:DS_thermalization_details}.

To understand the parametrics of the thermalization temperature, we note that each reaction-set involves an $\mathcal{O}(y'^2)$ decay-process, as well as a scattering process at either $\mathcal{O}(g'^4)$, $\mathcal{O}(g'^2 y'^2)$, or $\mathcal{O}(y'^4)$ in the DS couplings. Therefore, we can conservatively estimate $T_\text{th}$ by requiring that all processes are parametrically faster than Hubble expansion. Using Eq.~\eqref{eq:Y_FI_final} the relevant comparison is
\begin{align}
     \, \text{min}\left\{ y'^2 \frac{M^2}{M_\text{Pl} T_\text{th}}\;,\; 10^{-4} \, ( e^4 \kappa^2) \, g'^4 \;,\; 10^{-4} \,( e^4 \kappa^2) \, y'^4 \;,\; 10^{-4} \, ( e^4 \kappa^2) \, g'^2 y'^2 \right\} \sim \frac{T_\text{th}^2}{M_\text{Pl}^2} \, ,
    \label{eq:thermalization_parametric_comparison}
\end{align}
where we have assumed that the thermally-averaged cross sections in the high-$T$ limit are given by $\langle \sigma v \rangle \propto 1/T^2$, scaled by the appropriate couplings. Note that Eq.~\eqref{eq:thermalization_parametric_comparison} is equivalent to
\begin{align}
    T_\text{th} \sim M_\text{Pl} ~ \text{min}\{ y'^{2/3}\; , \; 10^{-2} \, e^2 \kappa \,  g'^2\;,\; 10^{-2} \, e^2 \kappa \,  y'^2\;,\; 10^{-2} \, e^2 \kappa \, g' y' \} \, .
    \label{eq:T_th}
\end{align}

\subsection{Dark Sector Energy Density Evolution}
\label{subsec:DS_temperature_evolution}

\begin{figure}
    \centering
    \includegraphics[width=0.75\linewidth]{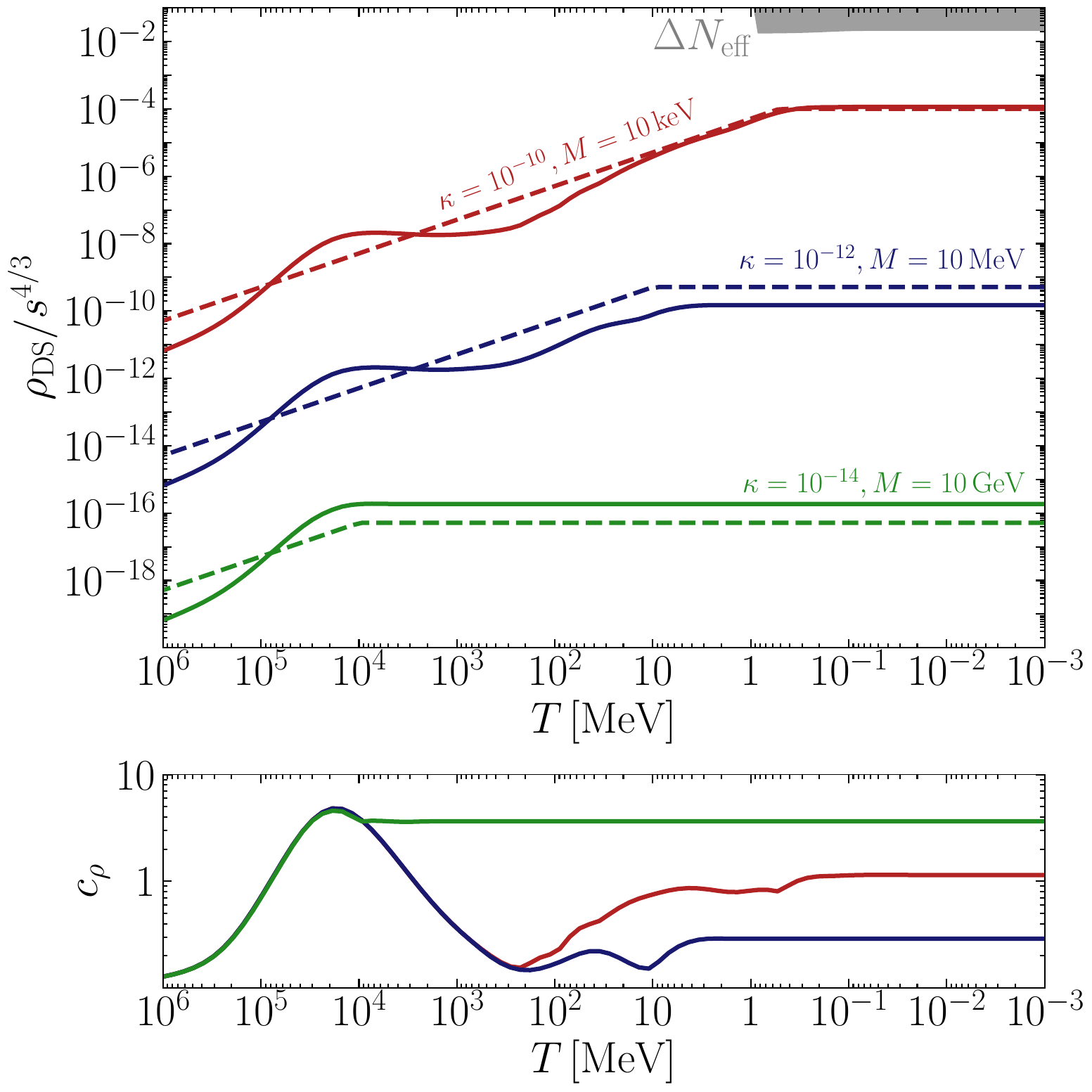}
    \caption{Evolution of the dark sector energy density for different choices of coupling, $\kappa$ (Eq.~\eqref{eq:kappa}), and masses for the dark charged particles, $M$, illustrated in different colors and labeled in the top panel. Solid lines are computed from the full numerical evaluation of Eq.~\eqref{eq:energy_Boltzmann_equation}, and dashed lines are the parametric expectation (Eq.~\eqref{eq:rho_DS_evolution} with $c_\rho = 1$). The solid gray region, labeled $\Delta N_\text{eff}$, indicates when the dark sector energy density would be in tension with limits on $\Delta N_\text{eff}$, discussed in detail in Sec.~\ref{subsec:Neff_bound}. In the lower panel, we plot $c_\rho$, which is an $\mathcal{O}(1)$ coefficient determining the ratio of the parametric expectation of $\rho_\text{DS} / s^{4/3}$ to the full numerical solution.}
    \label{fig:xi_fig}
\end{figure}

Once the frozen-in density of DCPs grows to a sufficiently large value to enable these number-changing reactions, the DS thermalizes to a temperature $T' \ll T$. In this case, DS number densities do not follow the simple evolution described by Eqs.~\eqref{eq:Y_boltz_FI} and \eqref{eq:Y_FI_final}. Instead, it is more useful to track the evolution of the DS energy density, $\rho_\text{DS}$, since DS number-changing reactions conserve energy and $\rho_\text{DS}$ can be used to determine the evolution of $T'$, which will be discussed in Sec.~\ref{subsec:ds_particle_decouple}. Parametrically, we expect that once DS number-changing reactions become active, the frozen-in DS energy evolves as $\rho_\text{DS} / \rho_\text{SM} \sim \Gamma_{\text{SM} \to X} / H \propto e^4 \kappa^2 M_\text{Pl} / T$, similar to the initial evolution of $Y_X$. 

This can be confirmed by solving the following Boltzmann equation~\cite{Chu:2011be,Bhattiprolu:2023akk},
\begin{align}
    \frac{d (\rho_\text{DS} / s^{4/3})}{dT} & = - \frac{c}{s^{4/3} H T} \sum_{i = \gamma^*, e, \mu, \dots} \, \sum_{X = \psi, \phi} R^\rho_{i \rightarrow X} \, ,
    \label{eq:energy_Boltzmann_equation}
\end{align}
where $R^\rho_{i \rightarrow X}$ is the integrated \textit{energy density} collision term. For example, the contribution from electron-positron annihilation is $R^\rho_{e \rightarrow \psi}$,
\begin{align}
    R^\rho_{e \rightarrow \psi} = \frac{T}{16 \, (2 \pi)^5} \int_{s_\text{min}}^\infty \dd s \, s \, \sqrt{1 - \frac{4 m_e^2}{s}} \sqrt{1 - \frac{4 M^2}{s}} \, \mathscr{M}^2_{ee \rightarrow \psi\psi} \, K_2(\sqrt{s}/T)
    \, ,
    \label{eq:R_rho_example}
\end{align}
where $s_\text{min} = 4 \, \max\{ m_e^2, M^2 \}$, and $\mathscr{M}^2_{ee \rightarrow \psi\psi}$ was defined previously in Eq.~\eqref{eq:M_scr_ee_psi_psi}. A complete list of the matrix elements needed to find $R^\rho$ for all relevant processes is given in App.~\ref{app:integrated_density_collision_terms}. In the large $T$ limit, $d (\rho_\text{DS} / s^{4/3}) / dT \propto - M_\text{Pl} / T^2$. Therefore, early in its evolution, $\rho_\text{DS} / s^{4/3}\propto M_\text{Pl} / T$, which continues until the SM temperature is $T \sim \max\{ m_e, M \}$. 

We connect this parametric understanding of the DS energy density evolution to the numerical solution of Eq.~\eqref{eq:energy_Boltzmann_equation} by introducing an $\mathcal{O}(1)$ temperature-dependent coefficient, $c_\rho$, defined through
\begin{align}
    \frac{\rho_\text{DS}}{s^{4/3}} = 10^{-4} \, c_{\rho} \,  e^4 \kappa^2 \, \left( \frac{M_\text{Pl}}{\text{max}\{ T, M, m_e \}} \right) \, .
    \label{eq:rho_DS_evolution}
\end{align}
In Fig.~\ref{fig:xi_fig}, we compare the numerical solution for $\rho_\text{DS} / s^{4/3}$ to the parametric estimate (i.e., Eq.~\eqref{eq:rho_DS_evolution} with $c_\rho = 1$) for a few different values of DCP masses and couplings. This shows that the rough parametric estimate of $\rho_\text{DS}$ matches the numerical solution within an order of magnitude.

\subsection{Dark Sector Particles Decoupling}
\label{subsec:ds_particle_decouple}

We now discuss how each particle in the DS decouples from the dark thermal bath, which determines the DS species' final abundance. Our focus in this section is two-fold. First, we derive and provide parametric formulas for the DM relic abundance in order to understand the DS couplings required for $\chi$ to achieve the observed relic density, 
\begin{align}
    Y_\text{DM} = \frac{\Omega_\text{DM} \, \rho_c}{m_\chi \, s_0} \sim 4 \times 10^{-5} \left( \frac{10 \, \text{keV}}{m_\chi} \right) \, ,
    \label{eq:Y_DM_relic}
\end{align}
where $\Omega_\text{DM} h^2 \simeq 0.12$, $\rho_\text{c} / h^2 \simeq 2.78 \times 10^{11} \, M_\odot \,\text{Mpc}^{-3}$, and $s_0 \simeq 2.5 \times 10^3 \, \text{GeV}^3$~\cite{Tiesinga:2021myr}. Second, we show that the DCPs only constitute a small fraction of the total DM density in our parameter space of interest.

Let us begin by introducing some quantities that will be important for each of the following sections. We define the ratio of the DS temperature, $T'$, to the SM temperature as $\xi \equiv T' / T$. Immediately after the DS thermalizes, each of the DS species is at temperature $T'$ (once $\psi$, $\phi$, and $\chi$ eventually decouple from $A'$, $T'$ is strictly the temperature of $A'$). As described above in Sec.~\ref{subsec:thermalization}, DS thermalization occurs once the temperature of the SM bath approaches $T_\text{th}$. We can find the corresponding initial DS temperature, $T'_\text{th}$, assuming that thermalization occurs instantaneously and conserves energy. This corresponds to taking $\rho_\text{DS} \sim T_\text{th}^{' \, 4}$ in Eq.~\eqref{eq:rho_DS_evolution}, which gives 
\begin{align}
    T'_\text{th} \sim 10^{-1} \, \sqrt{e^2 \kappa} \, M_\text{Pl}^{1/4} \, T_\text{th}^{3/4} \, .
\end{align}
Requiring that the DS thermalizes as a relativistic bath, $T_\text{th}' \gtrsim M$, then implies that the DS couplings are restricted to
\begin{align}
    (e^2\kappa)^{2/3} \, \text{min}\{ 10^{-1} \, y'^{2/3}\; , \; 10^{-3} \, e^2 \kappa \, g'^2\;,\; 10^{-3} \, e^2 \kappa \, y'^2\;,\; 10^{-3} \, e^2 \kappa \,  g' y' \} \gtrsim \left( \frac{M}{M_\text{Pl}} \right)^{4/3} \, .
    \label{eq:thermalization_condition}
\end{align}

The first particles to decouple from the DS bath will be the DCPs, due to their larger mass. This occurs shortly after $T'$ drops below $M$. Since this is an important temperature in all of the following cosmological scenarios, we define the SM temperature at which $T' = M$ as
\begin{align}
    T_* \equiv T(T' = M) =  \frac{M}{\xi(T_*)} \, .
    \label{eq:T_star_DCP}
\end{align}
The rate for processes involving only $\chi$ and $A'$ that change particle number (such as $\chi A' A' \leftrightarrow \chi A'$ and $A' A' A' A' \leftrightarrow A' A'$ ) are suppressed by the fact that they arise only at loop-level in the form of mass-dimension nine and twelve operators, respectively. As a result, once $T \lesssim T_*$ and DCPs decouple, DS number-changing reactions become inactive. This decoupling can occur before ($T_* > m_e$) or after ($T_* < m_e$) the SM is done injecting energy into the DS. As we will see, the evolution of the DCP and DM abundances, $Y_X$ and $Y_\chi$, respectively, strongly depend on this ordering, as shown schematically in Figs.~\ref{fig:scenario_I_yield} and \ref{fig:scenario_II_yield}. To address this dramatic difference in phenomenology, we will split our discussion into different sections accordingly. Sec.~\ref{subsubsec:scenario_I} corresponds to the scenario where $T_* < m_e$, while Sec.~\ref{subsubsec:scenario_II} focuses on $T_* > m_e$. 

\begin{figure}[ht!]
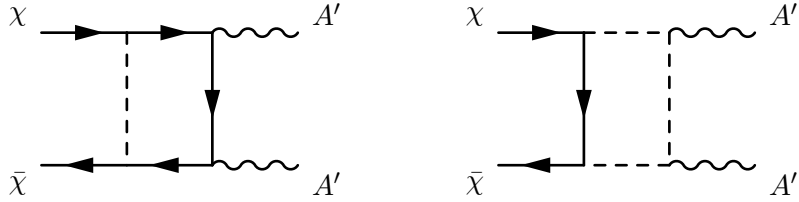

    \centering
    \include{feynman_diagrams/chi_AA_example_diagrams}
    \vspace{-1em}
    \caption{Two of the seven one-loop Feynman diagrams that contribute to $\chi$'s coupling to the dark photon, $A'$.}
    \label{fig:chi_AA}
\end{figure}
 
After the DCPs decouple, the DS bath consists solely of $\chi$ and $A'$. Although number-changing processes remain inactive, lower order reactions that interchange $\chi$ with $A'$ (such as $\chi \chi \leftrightarrow A' A'$ and $\chi A' \leftrightarrow \chi A'$) are sufficiently strong to maintain chemical and kinetic equilibrium between the two species. Since the DM is not charged under $U(1)'$, these interactions are generated radiatively, as shown in Fig.~\ref{fig:chi_AA}. The comoving density of $\chi$ freezes out once such processes eventually decouple as well. In particular, $\chi$ decouples from $A'$ as a \textit{hot} or \textit{cold} relic, depending on whether $n_\chi \, \langle \sigma v \rangle'_{\chi \chi \rightarrow A'A'}$ is smaller or larger than Hubble at $T' \sim m_\chi$, respectively, where 
\begin{align}
    \langle \sigma v \rangle'_{\chi\chi \rightarrow A'A'} \approx \frac{1}{2^2} \frac{y'^4 g'^4}{768 \pi^5}\frac{m_\chi^4}{M^6}
    \label{eq:DS_chichi_ApAp}
\end{align}
is the thermally-averaged $\chi\chi \rightarrow A'A'$ annihilation cross section in the non-relativistic limit, and the factor of $2^2$ comes from averaging over the DM spin states. Above, we have introduced notation used throughout this work where a prime superscript next to the brackets  (or lack thereof) denotes that a thermally-averaged cross section is evaluated at DS temperature $T'$ (or SM temperature $T$).  

We now begin our discussion of the different cosmological scenarios that can occur for different hierarchies between $T_*$ and $m_e$.

\subsubsection{Scenario I: $T_* < m_e$}
\label{subsubsec:scenario_I}

\begin{figure}[t!]
    \centering
    \includegraphics[width=\linewidth]{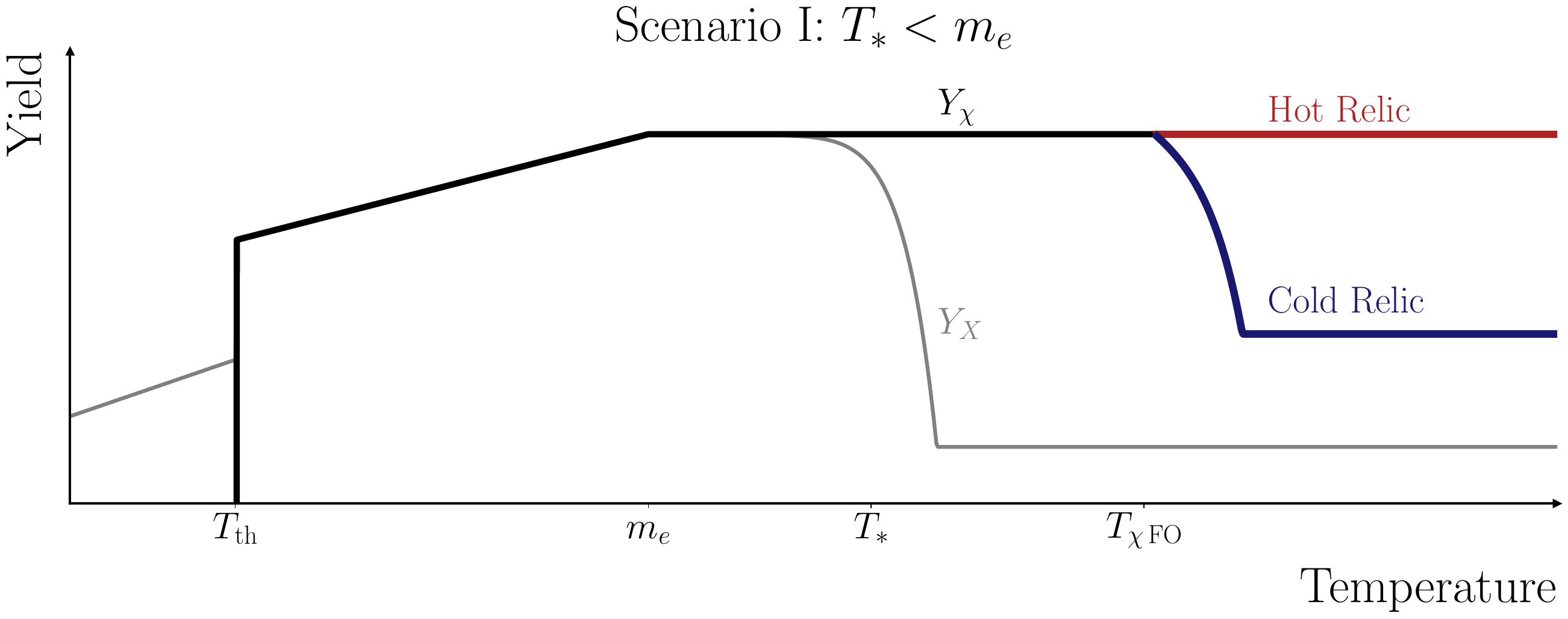}
    \caption{Evolution of the DS species yields in Scenario I, $T_* < m_e$, discussed in detail in Sec.~\ref{subsubsec:scenario_I}).}
    \label{fig:scenario_I_yield}
\end{figure}

In this first scenario, illustrated in Fig.~\ref{fig:scenario_I_yield}, the DCPs decouple from the DS \emph{after} SM-to-DCP freeze-in processes become negligible. As a result, the evolution of $T'$ is simple. For $T > m_e$, when SM-to-DS freeze-in is actively depositing energy into the DS, DS number-changing reactions are rapid, and therefore $T' \sim \rho_\text{DS}^{1/4}$, with $\rho_\text{DS}$ given by Eq.~\eqref{eq:rho_DS_evolution}. Later, when the SM temperature is $T < m_e$, the DS evolves as its own thermal bath independent of the SM, such that the temperature ratio $\xi$ is approximately fixed, up to small corrections from the evolving number of relativistic degrees of freedom. This gives
\begin{align}
    \xi \simeq \frac{1}{T}\left( \frac{30}{\pi^2 g_\text{DS}^*} \, \rho_\text{DS} \right)^{1/4} \sim 10^{-1} \, \sqrt{e^2\kappa} \, \left( \frac{M_\text{Pl}}{\text{max}\{ T, m_e \} } \right)^{1/4} \, ,
    \label{eq:xi_1}
\end{align}
where $g_{\text{DS}}^* = 2 + 4 \, (7/8) + 4 \, (7/8) + 2 = 7.5$ is the total effective relativistic degrees of freedom in the DS.

Given this behavior for $T'$, the DM relic abundance is then straightforwardly calculated depending on whether it decouples from $A'$ as a hot or cold relic (labeled ``HR" and ``CR," respectively). In particular, we find
\begin{align}
    Y_{\chi} \simeq \,  \begin{cases}
        \displaystyle \frac{90}{\pi^4 g_{*s}(m_e)} \xi_e^3  \sim 10^{-3} \, \left( e^2 \kappa \right)^{3/2} \left( \frac{M_\text{Pl}}{m_e} \right)^{3/4}  \,  & \text{\textcolor{BrickRed}{(HR)}} \\[1em]
        \displaystyle \frac{H}{s \langle \sigma v \rangle'_{\chi\chi \rightarrow A'A'}} \Bigg|_{T_{\chi\, \text{FO}}} \sim 10^6 \, \frac{\sqrt{e^2 \kappa}}{y'^4 g'^4} \left( \frac{M_\text{Pl}}{m_e} \right)^{1/4} \left( \frac{m_\chi}{M_\text{Pl}} \right) \left( \frac{M}{m_\chi} \right)^6 &  \text{\textcolor{MidnightBlue}{(CR)}} \, ,
    \end{cases}
    \label{eq:Y_DM_scenario_I}
\end{align}
where $\xi_e \equiv \xi(m_e)$, and $g_{*s}$ is the entropic effective degrees of freedom~\cite{Husdal:2016haj,Hooper:2024avz}. In the second line of Eq.~\eqref{eq:Y_DM_scenario_I}, we have estimated the freeze-out abundance of a cold relic by evaluating $H / s \langle \sigma v \rangle'$ at the SM temperature when $\chi \chi \leftrightarrow A' A'$ decouples, $T_{\chi\,\text{FO}} \sim m_\chi / \xi_e / 10$. This corresponds to the standard parametric estimate for DM that freezes out non-relativistically, and ignores logarithmic corrections dependent on the $\chi \chi \leftrightarrow A' A'$ interaction cross section~\cite{Hooper:2024avz} (such logarithmic corrections will be similarly omitted when reporting other freeze-out temperature parametrics below).

DCPs make up a small subcomponent of the total DM density, and their final abundance can be estimated in a similar manner. In particular, they generically decouple from the rest of the DS while non-relativistic, since they are directly charged under $U(1)'$. Hence, their abundance is given by an equation analogous to the second line of Eq.~\eqref{eq:Y_DM_scenario_I}, but involving instead the DCP annihilation processes to other lighter DS states (denoted as $X X \to \text{DS}$),
\begin{align}
    Y_X \simeq \frac{H}{s \, \langle \sigma v \rangle'_{X X \to \text{DS}}} \Bigg|_{T_{X \, \text{FO}}}
    \sim \frac{\sqrt{e^2 \kappa}}{\max\{ y'^4, g'^4\}} \left( \frac{M_\text{Pl}}{m_e} \right)^{1/4} \left( \frac{M}{M_\text{Pl}} \right) \, .
    \label{eq:relic_psi_light}
\end{align}
Analagous to Eq.~\eqref{eq:Y_DM_scenario_I}, we have evaluated the above expression at the SM temperature at which DCP freeze-out occurs, $T_{X \, \text{FO}} \sim T_* / 10 \sim M / \xi_e / 10$.

\subsubsection{Scenario II: $T_* > m_e$}
\label{subsubsec:scenario_II}

\begin{figure}[t]
    \centering
    \includegraphics[width=\linewidth]{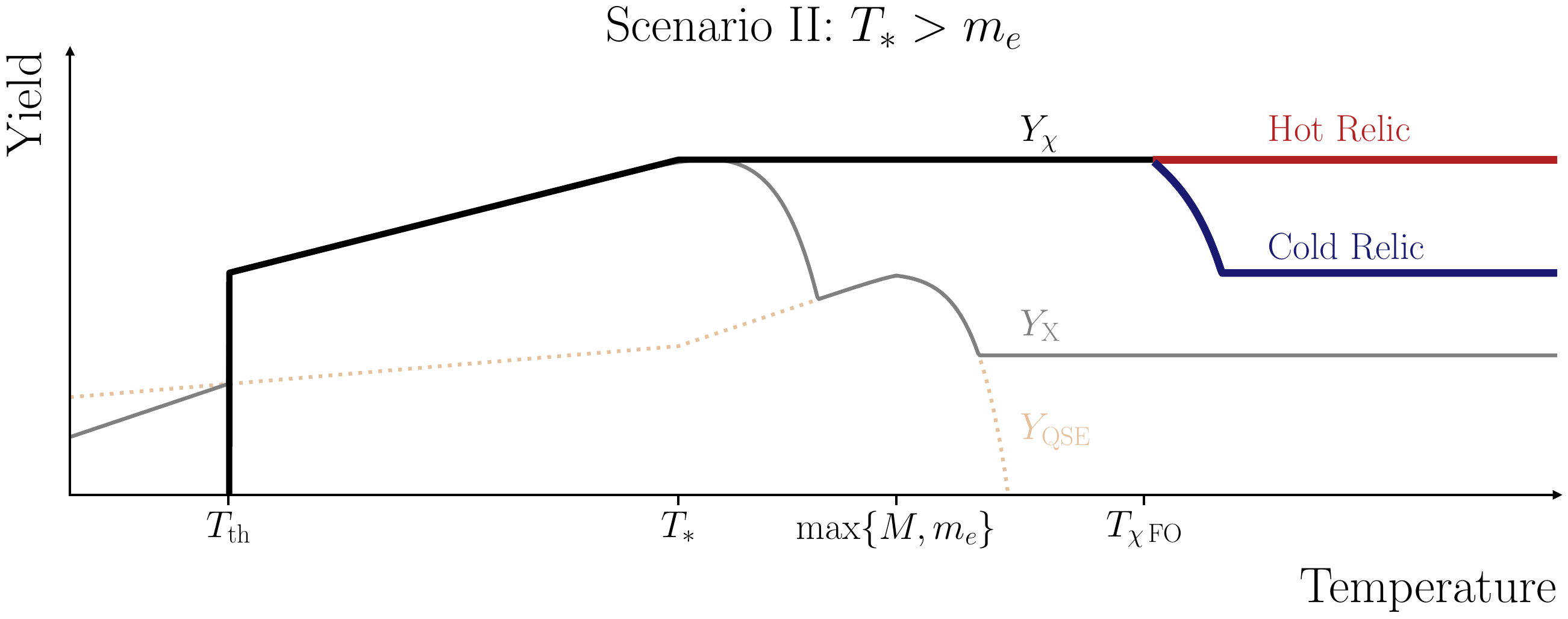}
    \caption{Evolution of the DS species yields in Scenario II, $T_* > m_e$, discussed in detail in Sec.~\ref{subsubsec:scenario_II}).}
    \label{fig:scenario_II_yield}
\end{figure}

In the second scenario, illustrated in Fig.~\ref{fig:scenario_II_yield}, the DCPs begin to decouple from the DS bath \emph{before} the SM is done injecting energy. In this case, the DS energy density $\rho_\text{DS}$ evolution is unmodified, such that we can use the previous results of Sec.~\ref{subsec:DS_temperature_evolution}. However, as we discuss below, the evolution of the DS temperature and number densities is significantly altered. 

We begin by describing the yields and temperature while $\chi$ is relativistic, $T' \gtrsim m_\chi$, before it decouples from the rest of the DS. Well before the DCPs decouple ($T > T_*$), the DS evolves as in Sec.~\ref{subsubsec:scenario_I}, corresponding to a thermal bath with zero chemical, $T' \sim \rho_\text{DS}^{1/4}$. At $T \sim T_*$, Eq.~\eqref{eq:xi_1} implies that
\begin{equation}
\xi (T_*) \sim 3 \times 10^{-2} \, (e^2\kappa)^{2/3} \, \left( \frac{M_\text{Pl}}{M} \right)^{1/3} \, ,
\label{eq:xiTstar}
\end{equation}
along with $Y_X (T_*) \sim Y_\chi (T_*) \sim Y_{A'} (T_*) \sim \xi (T_*)^3 \sim 10^{-4} \, (e^2 \kappa)^2 M_\text{Pl} / M$.

Later, once $T < T_*$, $\chi$ and $A'$ remain chemically and kinetically-coupled via 2-to-2 processes, but the absence of on-shell DCPs to facilitate 2-to-3 processes means that $\chi$ and $A'$ develop a common nonzero chemical potential. To estimate the DS temperature evolution during this later stage, we note that $A'$ is relativistic and always contributes at least an $\mathcal{O}(1)$ fraction of the total DS energy density. Hence, we take $T' \sim \rho_\text{DS} / n_{A'}$, where $\rho_\text{DS}$ is determined as in Sec.~\ref{subsec:DS_temperature_evolution} and the dark photon number density, $n_{A'}$, is determined below.

The behavior of $n_{A'}$ for $T < T_*$ depends directly on the DCP yield. The corresponding Boltzmann equation incorporating SM annihilations and DCP interactions with other DS species is given by
\begin{align}
    \frac{dY_X}{dT} & \approx -\frac{c}{s H T} \left[ \langle \sigma v \rangle_{\text{SM} \rightarrow XX} \left( n_e^\text{eq} \right)^2 - \langle \sigma v \rangle'_{X X \to \text{DS}} \left( n_X^2 - \left( n^{\prime\,\text{eq}}_X \right)^2 \right)  \right] \, .
    \label{eq:X_boltzmann_QSE_II}
\end{align}
Above, $\langle \sigma v \rangle_{\text{SM} \rightarrow XX} \left( n_e^\text{eq} \right)^2 \equiv \sum_{i = e,\mu, \ldots} n_i^2 \langle \sigma v \rangle_{ii \rightarrow XX}$, and $n_X^{\prime \, \text{eq}}$ is the DCP equilibrium number density evaluated at $T'$.\footnote{When $M > m_e$, it is more convenient to rewrite Eq.~\eqref{eq:X_boltzmann_QSE_II} as $( n_e^\text{eq})^2 \langle \sigma v \rangle_{\text{SM} \rightarrow XX} \rightarrow (n_X^\text{eq})^2\langle \sigma v \rangle_{XX \rightarrow \text{SM}}$, since $\langle \sigma v\rangle_{XX \rightarrow \text{SM}}$ does not exponentially decrease for $T < M$. However for simplicity we will use Eq.~\eqref{eq:X_boltzmann_QSE_II}.} When the SM temperature is $\text{max}\{ m_e , M \} \lesssim T \lesssim T_*$, the last term of Eq.~\eqref{eq:X_boltzmann_QSE_II} is negligible compared to the first, since $n^{\prime\,\text{eq}}_X$ is Boltzmann suppressed. In this limit, Eq.~\eqref{eq:X_boltzmann_QSE_II} simplifies to
\begin{align}
    \frac{dY_X}{dT} & \approx -\frac{c s}{H T} \langle \sigma v \rangle'_{X X \to \text{DS}} \left[ \left( Y_X^{\text{QSE}} \right)^2 - Y_X^2  \right]
    \, ,
    \label{eq:QSE_boltzmann_II}
\end{align}
where we have defined the density at \textit{quasi-static equilibrium} (QSE),
\begin{equation}
Y_X^\text{QSE} = \sqrt{\frac{\langle \sigma v  \rangle_{\text{SM} \rightarrow XX}}{\langle \sigma v  \rangle'_{X X \to \text{DS}}}} ~  \, \frac{n_e^\text{eq}}{s} \equiv \frac{n_\text{QSE}}{s}
\, ,
\label{eq:YQSE}
\end{equation}
corresponding to the yield that is achieved when SM annihilations to DCPs are balanced by DCP annihilations to other DS species. Therefore, provided that $n_\text{QSE}(T) \, \langle \sigma v \rangle'_{X X \to \text{DS}} \gg H$, the DCPs maintain QSE, $Y_X \simeq Y_X^\text{QSE}$ (see, e.g., Refs.~\cite{Chu:2011be,Bhattiprolu:2023akk,Bhattiprolu:2024dmh} for more detailed discussion regarding QSE). In terms of DS couplings, such QSE is maintained provided that $e^2 \kappa \max\{ y'^2, g'^2 \} \gtrsim M /M_\text{Pl}$, which is satisfied for our parameter space of interest.

This QSE population of DCPs sources a density of $\chi$ and $A'$ particles through, e.g., annihilation processes $X X \to \text{DS}$. Since $T \lesssim T_*$, corresponding to $T' \ll M$, we can ignore the inverse process $\text{DS} \to X X$, such that the resulting yield of relativistic $\chi$ and $A'$ particles is parametrically
\begin{equation}
Y_\chi^\text{QSE}  \sim Y_{A'}^\text{QSE} \sim \frac{n_\text{QSE} \, \langle \sigma v \rangle'_{X X \to \text{DS}}}{H} ~ Y_X^\text{QSE} \sim \frac{n_e^\text{eq} \, \langle \sigma v \rangle_{\text{SM} \to XX}}{H}
\, ,
\label{eq:YApQSE}
\end{equation}
where in the last equality we used Eq.~\eqref{eq:YQSE}.\footnote{Eq.~\eqref{eq:YApQSE} can be seen directly from the Boltzmann equation for $\chi$, where the backreaction is kinematically suppressed since $T' < M$, $-(H T / cs) \, dY_\chi / dT \approx \langle \sigma v \rangle_{\text{DS} \rightarrow \chi\chi}' n_\text{QSE}^2 \sim \langle \sigma v \rangle_{\text{SM} \rightarrow XX} \, (n_e^\text{eq})^2$ and we have assumed $\langle \sigma v \rangle'_{\text{DS} \rightarrow \chi\chi} \sim \langle \sigma v \rangle'_{X X \to \text{DS}}$.}  Hence, for $T \lesssim T_*$, the relativistic densities are $Y_\chi^\text{QSE} \sim Y_{A'}^\text{QSE} \sim 10^{-4} \, (e^2 \kappa)^2 M_\text{Pl} / \text{max}\{T ,  M, m_e \}$. This is comparable or smaller than the number density of particles sourced during $T \gtrsim T_*$ (see below Eq.~\eqref{eq:xiTstar}). 

Thus, the relativistic number density of $\chi$ and $A'$ particles sourced during $T \gtrsim T_*$ dominates over the density of those sourced later during $T \lesssim T_*$. However, it is this later population that dominates the total energy density, since it is created after DS number-changing reactions decouple, such that the resulting $\chi$ or $A'$ particles retain an average energy comparable to the SM temperature. Hence, for $T < T_*$, with $Y_{A'}$ given by the result below Eq.~\eqref{eq:YApQSE}, the DS temperature is given by
\begin{equation}
\xi (T < T_*) \simeq \frac{1}{3 T} \, \frac{\rho_\text{DS}}{n_{A'}}
\sim 10^{-1} \frac{M}{ \text{max}\{T ,  M, m_e \}}
\, .
\end{equation}

To summarize, the DS temperature evolution for $T_* > m_e$ is approximately given by
\begin{align}
    \xi \simeq 
    \begin{cases} 
        \displaystyle \frac{1}{T}\left( \frac{30}{\pi^2 g_\text{DS}^*}\rho_\text{DS} \right)^{1/4} \sim 10^{-1}\sqrt{e^2\kappa} \left( \frac{M_\text{Pl}}{T} \right)^{1/4} & (T > T_*) \\[3ex]
         \displaystyle \frac{1}{3 \, T}\frac{ \rho_\text{DS} }{ n_{A'} } \sim 10^{-1}\frac{M}{ \text{max}\{ T, M, m_e \} } & (T < T_*) \, .
    \end{cases}
    \label{eq:xi_2}
\end{align}
Note that since $\rho_\text{DS}$ increases until $T \sim \text{max}\{ M, m_e \}$ (Eq.~\eqref{eq:energy_Boltzmann_equation}), even after the DCPs decouple from the DS at $T \sim T_*$, the DS temperature continues to increase. This is because during the QSE evolution of the DCPs, their residual annihilations into $\chi$ and $A'$ are enough to increase $\xi$. In particular, this number density of $\chi$ and $A'$ particles is small, but they are created with an energy comparable to the SM temperature, which is much more energetic than the  DS particles sourced during $T > T_*$.

Although late-time QSE dynamics give rise to a subdominant yield of relativistic DCPs, it is this population that controls the non-relativistic freeze-out of DCPs. This can be gleaned from Eq.~\eqref{eq:YQSE}, which shows that the Boltzmann suppression of $Y_X^\text{QSE}$ occurs only after $T_{X \, \text{QSE FO}} \sim \text{max} \{ M, m_e\} / 10 \ll T_* / 10$. Hence, the standard estimate for the cold freeze-out density of DCPs is
\begin{align}
    Y_X \simeq \frac{H}{s \, \langle \sigma v \rangle'_{X X \to \text{DS}}} \Bigg|_{T_{X \, \text{QSE FO}}}
    \sim \frac{1}{\max\{ y'^4, g'^4\}} \left( \frac{M}{M_\text{Pl}} \right) \left( \frac{M}{\text{max}\{ M, m_e\}} \right) \, .
    \label{eq:Y_X_QSE_II}
\end{align}

Finally, to estimate the DM density, we note that $\chi$ decouples either relativistically or non-relativistically once the rate for $\chi \chi \to A' A'$ drops below the rate of Hubble expansion, as in the previous scenario. Using Eq.~\eqref{eq:xi_2}, the corresponding hot or cold relic DM abundances are then given by
\begin{align}
    Y_\chi \simeq \,  \begin{cases}
        \displaystyle \frac{90}{\pi^4 g_{*s}(T_*)} \xi_*^3  \sim 10^{-4} \, \left( e^2 \kappa \right)^2 \left( \frac{M_\text{Pl}}{M} \right) \,  & \text{\textcolor{BrickRed}{(HR)}} \\[1em]
        \displaystyle \frac{H}{s \langle \sigma v \rangle'_{\chi\chi \rightarrow A'A'}} \Bigg|_{\chi\, \text{FO}} \sim \frac{10^5}{y'^4 g'^4} \left( \frac{m_\chi}{M_\text{Pl}} \right) \left( \frac{M}{m_\chi} \right)^6 \left( \frac{M}{\text{max}\{M, m_e\}} \right) &  \text{\textcolor{MidnightBlue}{(CR)}} \, ,
    \end{cases}
    \label{eq:Y_DM_scenario_II}
\end{align}
where we used the shorthand $\xi_* \equiv \xi(T_*)$, and the SM temperature at DM freeze-out in the cold-relic-scenario is $T_{\chi \, \text{FO}} \sim m_\chi / ( 10 \, \xi_\chi)$, with $\xi_\chi \sim M / \text{max}\{ M, m_e \}$ (Eq.~\eqref{eq:xi_2}).

\section{Cosmological and Astrophysical Limits}
\label{sec:astrophysics}

We now consider a variety of cosmological and astrophysical limits on the model discussed in Sec.~\ref{sec:model}. In particular, we discuss $\Delta N_\text{eff}$ bounds in Sec.~\ref{subsec:Neff_bound}, DM self-interaction bounds in Sec.~\ref{subsec:SIDM}, limits from the formation of structure in Sec.~\ref{subsec:structure_formation}, and lastly limits from stellar energy loss in Sec.~\ref{subsec:stellar_energy_loss}.

\subsection{$\Delta N_\text{eff}$ Bound}
\label{subsec:Neff_bound}

The DS-SM couplings cannot be made arbitrarily large, otherwise too much energy is injected into the DS which would be in conflict with bounds on additional radiation in the early universe. This is parametrized by the change to the effective number of neutrinos,  $\Delta N_\text{eff}$, 
\begin{align}
    \frac{\rho_\text{DS}}{\rho} = \frac{7}{8} \left( \frac{4}{11} \right)^{4/3} \Delta N_\text{eff}
    \, .
    \label{eq:Neff_bound}
\end{align}
For concreteness, we evaluate Eq.~\eqref{eq:Neff_bound} at recombination, $T \sim 0.1 \, \text{eV}$, since the most stringent limits on $\Delta N_\text{eff}$ are from ACT~\cite{ACT:2025tim} and Planck~\cite{Planck:2018vyg} CMB measurements which, when combined with BAO measurements, limit $\Delta N_\text{eff} \lesssim 0.17$. We show the resulting bound on $\rho_\text{DS}$ as a gray shaded region in Fig.~\ref{fig:xi_fig}. In Sec.~\ref{sec:target_parameter_space}, we will also use this limit to place an upper bound on the coupling $\kappa$.

\subsection{Self-Interaction Bounds}
\label{subsec:SIDM}

DM cannot interact too strongly with itself, otherwise it would measurably alter observed structure on galactic and extragalactic scales. This is typically quoted as a limit on the transfer cross section for $\chi \chi \to \chi \chi$, $\sigma_\text{T} / m_\chi \lesssim 1 \, \text{cm}^2 / \text{g}$~\cite{ODonnell:2025pkw}. The corresponding self-interaction differential cross section is given by
\begin{align}
    \frac{d\sigma}{d\Omega} & = \frac{1}{4} \sum_\text{spins} \frac{|\mathcal{M}|^2}{64 \pi^2 E_\text{cm}^2} \approx \frac{1}{1024 \pi^2 \, m_\chi^2} \sum_\text{spins} |\mathcal{M}|^2 \, ,
\end{align}
where the factor of $1/4$ comes from averaging over the initial DM spin states, and $E_\text{cm} \approx 2 m_\chi$ in the non-relativistic limit and center of mass frame. When $d\sigma / d\Omega$ is independent of $\theta$, the transfer cross section is,
\begin{align}
    \sigma_\text{T} & \equiv \int d \Omega \, (1 - \cos{\theta}) \frac{d \sigma}{d\Omega} \approx 4 \pi \times \frac{d\sigma}{d\Omega} = \frac{1}{256 \, \pi \, m_\chi^2} \sum_\text{spins} |\mathcal{M}|^2 \, .
\end{align}

\begin{figure}
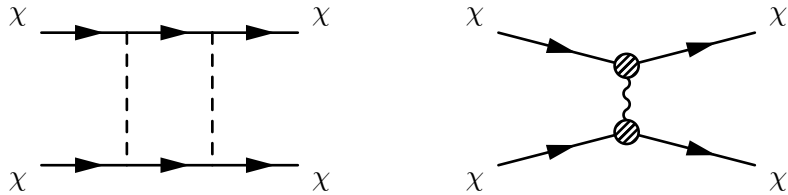

    \centering
    \include{feynman_diagrams/SIDM}
    \vspace{-1em}
    \caption{\textbf{Left:} One of the four Feynman diagrams at one loop contributing to $\chi$-$\chi$ Yukawa self-interactions. \textbf{Right:} One of the two Feynman diagrams contributing to magnetic dipole self-interactions. The blobs represent the MDM interaction between $\chi$ and $A'$ (Eq.~\eqref{eq:L_EFT}).}
    \label{fig:SIDM}
\end{figure}

There are a couple of interactions that can contribute to DM self-scattering at low-energies:

\textbf{Yukawa scattering}. The Yukawa interaction, $\mathcal{L} \supset \phi \bar{\psi} \chi + \text{h.c.}$, induces a loop-suppressed $\chi$-$\chi$ self-interaction, as shown in the left panel of Fig.~\ref{fig:SIDM}. When this dominates, the transfer cross section for DM self-interactions is
\begin{align}
    \sigma_\text{T} = \frac{y'^8}{18 \, ( 4\pi)^5} \frac{m_\chi^2}{M^4} \, .
    \label{eq:yukawa_sigma_T}
\end{align}
Therefore, to satisfy bounds on self-interactions, the DM must be heavier than
\begin{align}
    m_\chi > 0.3 \, \text{MeV} \times y^{\prime \, 8/3} \, \left( \frac{\sigma_\text{T} / m_\chi}{1 \, \text{cm}^2/\text{g} } \right)^{1/3} \left( \frac{M}{m_\chi} \right)^{-4/3} \, .
    \label{eq:m_chi_SIDM}
\end{align}
This shows that self-interaction bounds on the Yukawa interaction only limit the lightest DM masses when the Yukawa coupling is large and $M / m_\chi$ is small.
        
\textbf{Magnetic dipole scattering}. In addition to the Yukawa interaction, the DM can interact via the magnetic dipole interaction and the exchange of a light $A'$, as shown in the right panel of Fig.~\ref{fig:SIDM}. Typically, the limits on any self-interactions due to a light mediator are quite stringent, due to the small momentum transfer~\cite{Knapen:2017xzo}. However, higher-dimension operators typically introduce powers of momentum at the effective vertices, which can cancel those coming from from a long-ranged propagator. This is precisely what happens here for the MDM operator in Eq.~\eqref{eq:L_EFT}. We find that when this dominates DM self-interactions, the transfer cross section is
\begin{align}
    \sigma_\text{T} = \frac{3 m_\chi^2}{4 \pi} \mu_\chi'^4 = \frac{3 g'^4 y'^8}{4 (32)^4 \pi^9} \frac{m_\chi^2}{M^4} \, ,
\end{align}
which is not enhanced by the DM's small velocity and is always much smaller than Eq.~\eqref{eq:yukawa_sigma_T}.

\subsection{Structure Formation}
\label{subsec:structure_formation}

Light DM that is part of a thermalized DS can also suppress the observed matter power spectrum at small scales through DM free-streaming and dark acoustic oscillations~\cite{Bertoni:2014mva,Berlin:2018ztp}. The former is relevant to many models of light DM, since its characteristic velocity is enhanced for smaller masses, suppressing the growth of structure below the corresponding free-streaming scale. The latter effect is relevant for DM that couples to dark radiation, which collisionally imparts pressure to the DM, slowing the growth of its perturbations until such processes decouple. 

These processes dampen the matter power spectrum on length scales below a characteristic comoving scale $\lambda_\text{cut} = \text{max}\{ \lambda_\text{ao},  \lambda_\text{fs} \}$, where $\lambda_\text{ao}$ and $\lambda_\text{fs}$ incorporate effects from dark acoustic oscillations and free-streaming, respectively. These comoving scales are calculated as
\begin{align}
    \lambda_\text{ao} & = \int^{t_\text{kd}}_0 \frac{\dd t}{a} = \frac{1}{a_\text{kd} H_\text{kd}} \label{eq:lambda_ao}\\
    \lambda_\text{fs} & = \int_{t_\text{kd}}^{t_\text{eq}} v_\chi \frac{\dd t}{a} \simeq \lambda_\text{ao} \frac{p_\text{kd}}{m_\chi} \log{ \left( \frac{a_\text{eq}}{a_\text{kd}} \right) } \, , \label{eq:lambda_fs}
\end{align}
where quantities with the subscript ``kd" are evaluated at $T_\text{kd}$, the SM temperature at which $\chi$ kinetically decouples from the $A'$ population. We assume kinetic decoupling happens at a DS temperature of $T'_\text{kd} \ll m_\chi$, such that $T'_\text{kd} \ll p_\text{kd} \ll m_\chi$ and the DM velocity is $v_\chi \approx p_\chi / m_\chi = p_\text{kd} (a_\text{kd} / a) / m_\chi$, where $p_\chi$ is the typical $\chi$ momentum after kinetic decoupling, $p_\text{kd}$ is $p_\chi$ at kinetic decoupling, and we have simplified these expressions assuming radiation domination, $H \approx H_\text{kd} (a_\text{kd} / a)^2$. Comparing Eq.~\eqref{eq:lambda_ao} to Eq.~\eqref{eq:lambda_fs}, we find that $\lambda_\text{ao} \gg \lambda_\text{fs}$ and hence dark acoustic oscillations are generally more constraining.

To compute the comoving length scales in Eqs.~\eqref{eq:lambda_ao} and~\eqref{eq:lambda_fs}, one must find $T_\text{kd}$, the SM temperature at which $\chi$ and $A'$ kinetically decouple. Kinetic equilibrium is maintained via the efficiency of $\chi A^\prime \leftrightarrow \chi A^\prime$ in transferring momentum~\cite{Profumo:2025uvx}. On average, a typical scatter transfers momentum $\sim T^\prime$. Hence, the total momentum transferred to the DM after $N$ scatters is roughly $\sqrt{N} \, T^\prime$, which is comparable to the DM momentum $\sim \sqrt{m_\chi \, T^\prime}$ (assuming $T' \ll m_\chi$) after $N \sim m_\chi / T^\prime \gg 1$ scatters. Thus, $\chi$ and $A^\prime$ kinetically decouple at temperature $T_\text{kd}$ when $\gamma \sim H$, where the effective interaction rate $\gamma$ is given by
\be
\gamma \sim \frac{T^\prime}{m_\chi} \, n_{A^\prime} \, \langle \sigma v \rangle_{\chi A^\prime \to \chi A^\prime}
\label{eq:kindecgamma}
~.
\ee
To evaluate the right-hand side, we assume that $\langle \sigma v \rangle_{\chi A' \rightarrow \chi A'} \approx \langle \sigma v \rangle_{\chi \chi \rightarrow A'A'}$ given in Eq.~\eqref{eq:DS_chichi_ApAp}.

Damping of structure formation below $\lambda_\text{cutoff}$ means that there is a minimum mass for gravitationally collapsed DM structures, 
\be
M_\text{cut} \sim \frac{4 \pi}{3} \, \Omega_\text{DM} \, \rho_\text{c} \, \lambda_\text{cut}^3 \sim 1.4 \times 10^8 \ M_\odot \times \bigg( \frac{\lambda_\text{cut}}{0.1 \ \text{Mpc}}\bigg)^3
~.
\ee
Various astrophysical observations sensitive to the matter power spectrum on small scales, such as Milky Way satellite counts and the Lyman-$\alpha$ forest, exclude $M_\text{cut} \gtrsim (10^7 - 10^9) \ M_\odot$~\cite{Strigari:2008ib,Polisensky:2010rw,Vegetti:2012mc,Viel:2013fqw,Vegetti:2014lqa,Baur:2015jsy,Garzilli:2015iwa,Irsic:2017ixq,Kim:2017iwr}, which corresponds to requiring that $\lambda_\text{cut} \lesssim 0.1 \, \text{Mpc}$. Using Eq.~\eqref{eq:lambda_ao}, this can be translated into the requirement that $T_\text{kd} \gtrsim \text{keV}$. 

\subsection{Stellar Energy Loss}
\label{subsec:stellar_energy_loss}

DS particles can be produced in the cores of stellar objects and contribute to their energy loss, altering their observed evolutionary history. Therefore, regions of parameter space that predict excessive energy loss are excluded; these bounds are referred to as stellar cooling constraints. For DS particles with masses below approximately 10 keV, the strongest constraints come from red giants, while for heavier particles, the dominant bounds arise from Supernova 1987A. In this section, we consider constraints from both systems.

DCPs ($\psi$ and $\phi$) can be produced in the cores of stellar objects via kinetic mixing. Since we have assumed that the dark photon mass is negligibly small, the resulting constraints are equivalent to those on millicharged particles. Accordingly, we adopt the cooling bounds on millicharged particles from Supernova 1987A~\cite{Chang:2018rso} and red giants~\cite{Vogel:2013raa}. We note that the dark photon itself does not contribute to stellar cooling, since the kinetic mixing term for an on-shell dark photon is suppressed when the dark photon mass is small~\cite{Redondo:2008ec,An:2013yfc,Chang:2016ntp}.

DM particles ($\chi$) are produced in stellar cores through the dipole operator. Stellar constraints on dipole operators have been studied in Refs.~\cite{Chu:2019rok, Chang:2019xva}. However, these bounds are valid only in the regime $M \gg T_\text{core}$, where $M$ is the mass of the virtual particles that radiatively generate the dipole interaction and $T_\text{core}$ is the temperature of the stellar core. If the stellar core temperature exceeds $M$, the effective operator coefficient $\mu'_\chi$ in Eq.~\eqref{eq:L_EFT} no longer matches the expression in Eq.~\eqref{eq:dark_magnetic_dipole_moment}; in this case, the appropriate scale controlling the size of the interaction is $T_\text{core}$ instead of $M$. For $M \ll T_\text{core}$, the resulting bounds become independent of $M$, only depend on $\kappa$, and scale in the same way as those for millicharged particles. A detailed calculation is required to derive these bounds properly, which is beyond the scope of this work. In our analysis, we therefore apply the cooling bounds on the dipole operator only for $M > M_\textrm{max}$, where $M_\textrm{max}$ is the largest $M$ at which DCPs can be constrained by stellar objects. For $M < M_\textrm{max}$, we simply adopt the same coupling constraints as at $M = M_\textrm{max}$.

\section{Target Parameter Space For Direct Detection}
\label{sec:target_parameter_space}

Given the infrared freeze-in production mechanism discussed in Sec.~\ref{sec:cosmology}, and cosmological and astrophysical limits discussed in Sec.~\ref{sec:astrophysics}, we now begin our discussion of the detectability of this model from the perspective of direct detection. Since magnetic dipole DM interacts with a light kinetically-mixed dark photon, the light DM direct detection experiments sensitive to millicharge particles (e.g., those using electron ionization in noble liquids~\cite{XENON:2021qze,XENON:2022ltv,DarkSide:2022knj}, electronic excitations in crystal targets~\cite{CDEX:2019exx,CDEX:2022kcd,DAMIC-M:2023gxo,DAMIC-M:2025luv,EDELWEISS:2020fxc,SENSEI:2023zdf,SuperCDMS:2019jxx,SuperCDMS:2020ymb,Oscura:2022vmi}, or single phonon excitations~\cite{TESSERACT:2025tfw}) are also sensitive to this light DM candidate. Furthermore, magnon excitations are expected to be much more sensitive to light magnetic dipole DM~\cite{Trickle:2019ovy,Trickle:2020oki,Esposito:2022bnu,Marocco:2025eqw}, since they utilize the dominantly spin-dependent interactions at low energies~\cite{Krnjaic:2024bdd,Berlin:2025uka}.

\begin{figure}[t!]
    \centering
    \includegraphics[width=1.07\linewidth]{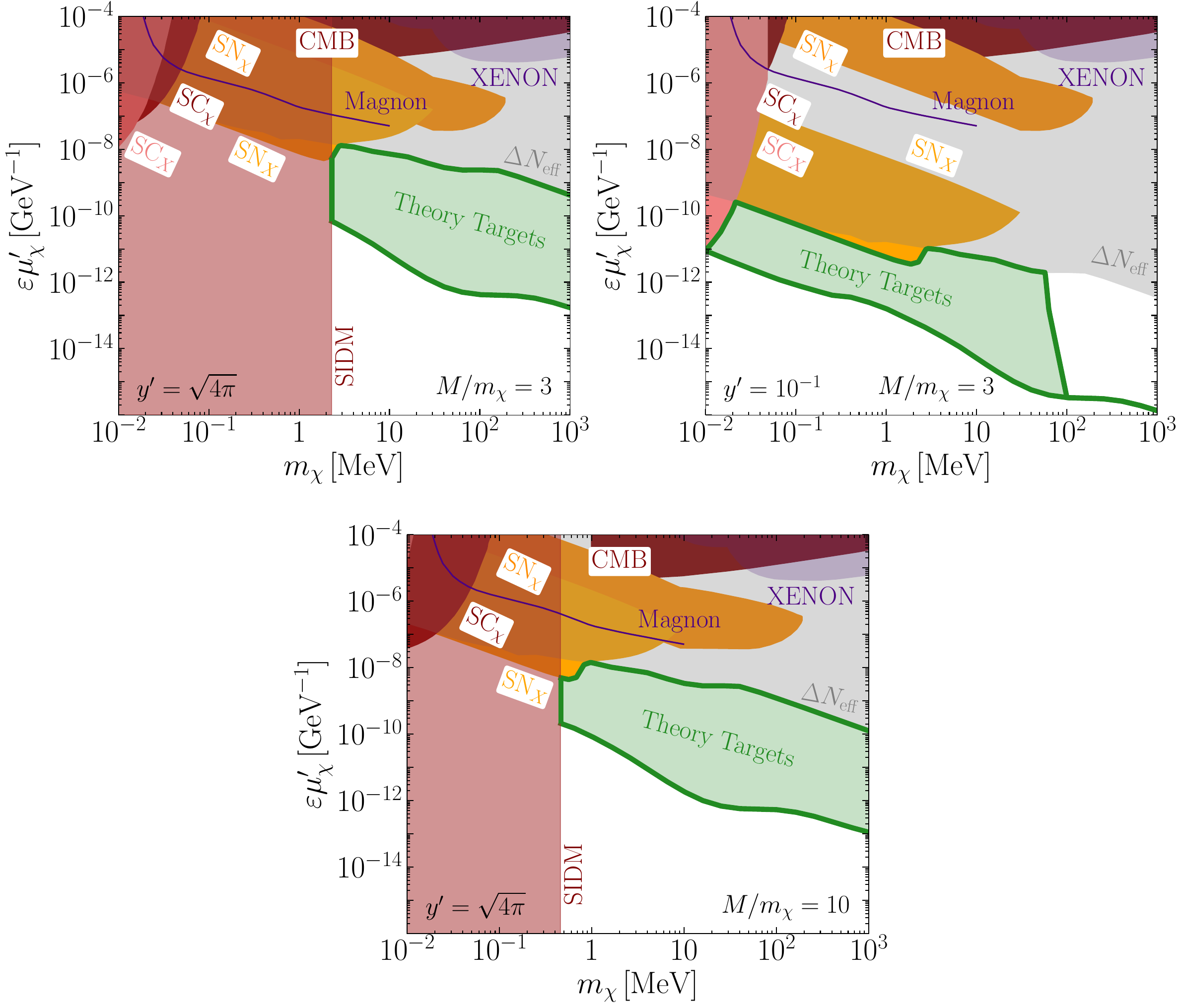}
    \caption{Overview of the available parameter space for light magnetic dipole dark matter (DM) that is produced from infrared freeze-in of dark charged particles (DCPs), shown in terms of the couplings most relevant for direct detection, $\varepsilon \mu_\chi'$, for different dark sector (DS) couplings and DCP-to-DM mass ratio combinations. In the parameter space outlined in green (labeled ``Theory Targets"), there exist viable DS couplings for which freeze-in produces the correct relic abundance of DM. The purple line (labeled ``Magnon") is the projected sensitivity of future magnon-based direct detection experiments~\cite{Trickle:2019ovy,Berlin:2025uka}, and the shaded purple region (labeled ``XENON") is excluded by the XENON10 and XENON1T experiments~\cite{Catena:2019gfa}. The gray region is excluded by limits from $\Delta N_\text{eff}$, as discussed in Sec.~\ref{subsec:Neff_bound}. The solid orange and red regions are excluded from stellar energy loss channels, as discussed in Sec.~\ref{subsec:stellar_energy_loss}; dark colors correspond to limits from production of the DM due to it's dark magnetic dipole moment (labeled ``$\text{SC}_\chi$" and ``$\text{SN}_\chi$" for RG stellar energy loss and Supernova energy loss, respectively~\cite{Chu:2018qrm,Chang:2019xva,Chu:2019rok}), and light colors correspond to limits from the production of the DCPs (labeled ``$\text{SC}_X$" and ``$\text{SN}_X$" for RG stellar energy loss and Supernova energy loss, respectively~\cite{Chu:2018qrm,Chang:2019xva,Chu:2019rok}). The solid maroon region (labeled ``CMB") corresponds to limits from CMB energy injection from annihilations to $e^+e^-$~\cite{Chu:2018qrm}. The light red region (labeled ``SIDM") corresponds to limits from self-interacting DM, as discussed in Sec.~\ref{subsec:SIDM}.}
    \label{fig:multi-panel}
\end{figure}

The parameter combination which determines the detectability of direct detection is $\varepsilon \mu_\chi'$, since $\varepsilon$ controls how strong the mediating dark photon couples to the electron, and $\mu_\chi'$ determines the DM's coupling to the dark photon (as shown in Eq.~\eqref{eq:L_EFT}).\footnote{As discussed in footnote~\ref{foot}, in the massless $A'$ limit, $\varepsilon \mu_\chi'$ is the $\chi$'s visible magnetic dipole moment.} A summary of our results is shown in Fig.~\ref{fig:multi-panel} for different choices of the DS Yukawa, $y'$, and DCP-to-DM mass ratio, $M / m_\chi$. The green regions labeled ``Theory Targets" are the viable and cosmologically-motivated parts of parameter space for light magnetic dipole DM. 
Notably, the projected sensitivity of the single magnon-based direct detection experiments in Refs.~\cite{Trickle:2019ovy,Trickle:2020oki,Berlin:2025uka}, as shown in Fig.~\ref{fig:multi-panel} (labeled ``Magnon"), is within an order of magnitude in probing these theory targets. Additionally, Ref.~\cite{Hochberg:2025rjs} has recently identified other magnetic targets which could further boost the sensitivity of single magnon-based experiments.

For the smallest couplings within this ``Theory Target" parameter space, light magnetic dipole DM freezes out as a hot relic (Sec.~\ref{subsec:ds_particle_decouple}) (i.e., $\chi$ chemically decouples from the $A'$ while relativistic). For the fixed values of $y'$ and $M / m_\chi$ in Fig.~\ref{fig:multi-panel}, this corresponds to small values of $g'$ since this minimizes the $\chi \chi \leftrightarrow A'A'$ interaction strength at late times (Eq.~\eqref{eq:DS_chichi_ApAp}). In this case, the DM relic abundance only depends on $\kappa = \varepsilon g' / e$. Therefore, for fixed $y'$,  $M / m_\chi$, and $m_\chi$, there is only a single value of $\varepsilon \mu_\chi'$ which satisfies the relic abundance. This is shown by the bottom of the outlined green region. Using the results of Sec.~\ref{sec:cosmology}, this corresponds to
\begin{align}
    (\varepsilon \mu_\chi')_\text{HR} \sim \begin{cases}
        \displaystyle 10^{-12} \, \text{GeV}^{-1} \, y'^2 \left( \frac{\text{MeV}}{m_\chi} \right)^{5/3} \left( \frac{M}{m_\chi} \right)^{-1} & \text{Scenario I} \\[2ex] 
        \displaystyle 10^{-11} \, \text{GeV}^{-1} \, y'^2 \left( \frac{\text{MeV}}{m_\chi} \right) \left( \frac{M}{m_\chi} \right)^{-1/2}  & \text{Scenario II} 
        \, ,
    \end{cases}
    \label{eq:hot_relic_couplings}
\end{align}
where Scenarios I and II are discussed in detail in Secs.~\ref{subsubsec:scenario_I} and~\ref{subsubsec:scenario_II}, respectively. Generally, Scenario I corresponds to DM much lighter than the electron, and Scenario II corresponds to DM much heavier than the electron.

For greater values of $\varepsilon \mu_\chi'$, the predicted DM relic abundance is too large unless $\chi$-$A'$ interactions are relevant. Indeed, if $g'$ is sufficiently large, the DM and dark photon can stay chemically coupled long enough to deplete the DM abundance through $\chi \chi \to A' A'$. This is the cold relic scenario discussed in Sec.~\ref{subsec:ds_particle_decouple}, which is a viable cosmology for values of $\varepsilon \mu_\chi'$ larger than those shown in Eq.~\eqref{eq:hot_relic_couplings}. However, for a fixed mass ratio $M / m_\chi$ and perturbative $g'$, there is a maximum DM mass for which the cold relic density can deplete down to the observed abundance today. This maximum mass can be understood parametrically by requiring that the cold relic abundance at $T_{\chi \, \text{FO}} \sim m_\chi / 10$ (Eq.~\eqref{eq:Y_DM_scenario_II}) agrees with the observed DM density, 
\begin{align}
    m_\chi \lesssim 200\, \text{GeV} \, y'^2 \left( \frac{M}{m_\chi} \right)^{-3}\, ,
    \label{eq:m_max_cold_relic}
\end{align}
where we demanded perturbativity in the form $g' \lesssim \sqrt{4 \pi}$. This is evident as the right edge of the viable parameter space in the top-right panel Fig.~\ref{fig:multi-panel}. This cutoff is not solely a function of $m_\chi$, since the exact freeze-out temperature for large DCP masses (Scenario II) depends logarithmically on $\varepsilon g'$.

The top-left edge of the viable parameter space of Fig.~\ref{fig:multi-panel} is set by the cosmological and astrophysical bounds discussed in Sec.~\ref{sec:astrophysics}. Of these, the limits that depend directly on the effective magnetic dipole $\varepsilon \mu_\chi'$ are the same in each panel (direct detection, $\text{SC}_\chi$ and $\text{SN}_\chi$). Since $\Delta N_\text{eff}$ depends only on the combination $\varepsilon g'$, limits from the effective number of neutrino species scale as $\propto y'^2 \, (m_\chi / M)$ between panels. Furthermore, the stellar energy loss limits for DCPs ($\text{SC}_X$ and $\text{SN}_X$) scale as $\propto y'^2$ for a fixed mass ratio; for larger $M / m_\chi$ mass ratios, these limits shift to smaller $m_\chi$ since they cease to apply when $M > T_\text{core}$ (Sec.~\ref{subsec:stellar_energy_loss}).

Not shown in Fig.~\ref{fig:multi-panel} are requirements for DS thermalization (Sec.~\ref{subsec:thermalization}) and limits from structure formation (Sec.~\ref{subsec:structure_formation}), since these depend on the specific value of $g'$, which can be chosen such that these considerations do not limit the available parameter space in Fig.~\ref{fig:multi-panel}. For instance, Eq.~\eqref{eq:thermalization_condition} implies that thermalization within the DS occurs provided that

\begin{align}
    (\varepsilon \mu_\chi')_\text{th} \gtrsim 10^{-14} \, \text{GeV}^{-1} \, y'^2 \left( \frac{10^{-2}}{g'} \right)^{6/5} \left( \frac{M}{m_\chi} \right)^{1/5} \left( \frac{\text{GeV}}{m_\chi} \right)^{1/5}
    \, ,
\end{align}
where we have assumed $y' \gg g'$ (appropriate for the left and middle panels in Fig.~\ref{fig:multi-panel}). This shows that for $g' \gtrsim 10^{-2}$, thermalization within the DS occurs over the ranges of interest for $m_\chi$ and $\varepsilon \mu_\chi'$.

Limits derived from structure formation (Sec.~\ref{subsec:structure_formation}) are important for large DS couplings and small DM masses (i.e., dominantly Scenario I of Sec.~\ref{subsubsec:scenario_I}), and therefore could be relevant for parts of parameter space where $\chi$ freezes out as a cold relic. In this case, it is straightforward to understand the parametrics governing $\chi$-$A'$ kinetic decoupling. Indeed, the DM-DS cross section which determines the abundance in  Eq.~\eqref{eq:DS_chichi_ApAp} also roughly determines the kinetic decoupling temperature via Eq.~\eqref{eq:kindecgamma}. The kinetic decoupling temperature can then be found parametrically to be 
\begin{align}
    T_\text{kd} \sim 10 \, \text{keV} \left( \frac{m_\chi}{10 \, \text{keV}} \right)^{1/2} \left( \frac{10^{-1}}{\xi} \right)^{5/2}
    \, ,
\end{align}
when $\chi$ constitutes a cold relic constituting all of the DM. From this we see that the impact of kinetic decoupling on structure formation is only relevant for $m_\chi \lesssim 10 \, \text{keV}$, since it is only then that $T_\text{kd} \gtrsim \text{keV}$ (see the discussion in Sec.~\ref{subsec:structure_formation}). Therefore, in Fig.~\ref{fig:multi-panel} we only show parameter space  for $m_\chi > 10 \, \text{keV}$. 

Lastly, we compare the theory targets of this work with other models with similar DS content. If $y'$ and $g'$ are not sufficiently large to allow the DS to self-thermalize, then the neutral species $\chi$ is not populated. In this case, the DCPs can constitute the DM, just as in the conventional millicharged scenario. The value of $\kappa$ required for this is given by $\kappa_\text{MCP} \sim 10^{-11} \, \left( \text{max}\{ M, m_e \} / M \right)^{1/2}$~\cite{Hall:2009bx}. We can compare this to the value of $\kappa$ required to generate magnetic dipole dark matter as a hot relic in Scenario I and II (Secs.~\ref{subsubsec:scenario_I} and~\ref{subsubsec:scenario_II}, respectively),  denoted as $\kappa_\text{HR}$, 
\begin{align}
    \frac{\kappa_\text{HR}}{\kappa_\text{MCP}} \sim \begin{cases}
        \displaystyle 10^{-1} \left( \frac{100 \, \text{keV}}{m_\chi} \right)^{2/3} \left( \frac{M}{m_e} \right)^{1/2}  & \text{Scenario I} \\[4ex] 
        \displaystyle \frac{M}{\sqrt{m_\chi \, \text{max}\{M, m_e \}} } & \text{Scenario II} \, ,
    \end{cases}
\end{align}
where $M < m_e$ in Scenario I. Additionally, we can compare the value of $\varepsilon \mu_\chi'$ required for infrared freeze-in, shown in Fig.~\ref{fig:multi-panel}, to that required from UV freeze-in of magnetic-dipole DM in Ref.~\cite{Chang:2019xva}. From Ref.~\cite{Chang:2019xva}, the value of $\varepsilon\mu'_\chi$ required by UV freeze-in  is $(\varepsilon \mu_\chi')_\text{UV} \sim 10^{-12} \, ( T_\text{RH} m_\chi )^{-1/2}$, where $T_\text{RH}$ is the reheat temperature. Again, comparing to our hot relic scenarios in Eq.~\eqref{eq:hot_relic_couplings}, we find
\begin{align}
    \frac{(\varepsilon \mu_\chi')_\text{HR}}{(\varepsilon \mu_\chi')_\text{UV}} \sim \,  \begin{cases}
        \displaystyle 10^{-2} \, y'^2 \left( \frac{\text{MeV}}{m_\chi} \right)^{7/6} \left( \frac{M}{m_\chi} \right)^{-1} \left( \frac{T_\text{RH}}{\text{GeV}} \right)^{1/2} & \text{Scenario I} \\[4ex] 
        \displaystyle 10^{-1} \, y'^2 \left( \frac{\text{MeV}}{m_\chi} \right)^{1/2} \left( \frac{M}{m_\chi} \right)^{-1/2} \left( \frac{T_\text{RH}}{\text{GeV}} \right)^{1/2} & \text{Scenario II} \, .
    \end{cases}
\end{align}

\section{Conclusions}
\label{sec:conclusions}

Freeze-in of keV-GeV mass dark matter (DM) due to an ultralight kinetically-mixed dark photon has long been an important benchmark model for direct detection experiments because of its predictability (the parameters determining the relic abundance also determine detectability), testability, and UV-insensitivity (assuming negligible earlier production through other means). The recent DAMIC results~\cite{DAMIC-M:2025luv} probing this model in the MeV-GeV mass range have generated new questions about what other light DM models such direct detection experiments are sensitive to, since DM couplings to the Standard Model (SM) smaller than the usual freeze-in benchmark typically lead to under-population of the DM. Therefore, understanding what other light DM models can be probed with these experiments, such as those with additional dark sector (DS) degrees of freedom, is an important model-building question today. This model-building challenge also presents an opportunity to uncover DM models with novel low-energy phenomenology that may best be probed with experiments other than those sensitive to electron excitations.

In this work, we identified a DM model whose low-energy phenomenology is determined by a dark magnetic dipole moment (MDM) that is generated by virtual dark charged particles (DCPs) (Sec.~\ref{sec:model}). We showed that such a model has a variety of UV-insensitive cosmological histories that emerge once DS thermalization occurs after a period of freeze-in (Sec.~\ref{sec:cosmology}), and we illustrated the parameter space allowed by cosmological and astrophysical bounds (Sec.~\ref{sec:astrophysics}) that is relevant for future direct detection experiments (Sec.~\ref{sec:target_parameter_space}). The allowed parameter space identified here (Fig.~\ref{fig:multi-panel}), specifically in the limit of relatively large DS couplings and small DCP-to-DM mass ratios, serves as a theoretically motivated benchmark target for future direct detection experiments. We find that a future single-magnon based experiment could offer the best opportunity to probe these models, due to their low-thresholds and sensitivity to dominantly spin-dependent interactions at low-energies.

There are other phases of the theory discussed in Sec.~\ref{sec:model} which may be phenomenologically interesting for DCPs heavier than an MeV. For example, with even larger DS couplings, $2 \rightarrow 3$ processes in the DS can remain active while the SM injects energy via the DCPs. This would increase the abundance before the DM decouples, necessitating smaller couplings for the hot relic scenario, yet raising the necessary DS couplings for a cold relic. Additionally, we have focused on the regime where the DS-SM coupling is small, and the DS self-couplings are large. A flipped hierarchy could allow the DCPs to thermalize with the SM, and then freeze-in the neutral DM particles from their interaction with the DCP. If the DCPs leave thermal equilibrium with the SM before BBN, this could be a way to weaken bounds from cosmological measurements of $\Delta N_\text{eff}$. Lastly, understanding current and future limits on magnetic dipole DM from semiconductor-based electron excitation experiments is important, and will require incorporating high-momentum effects from all-electron reconstruction~\cite{Griffin:2021znd,Krnjaic:2024bdd}, due to the dimension-five nature of the MDM interaction.

\acknowledgments

We would like to thank Prudhvi Bhattiprolu, Rouven Essig, Simon Knapen, Robert McGehee, Katelin Schutz, Jessie Shelton, and Mike Wagman for helpful discussion. This manuscript has been authored by FermiForward Discovery Group, LLC under Contract No. 89243024-CSC000002 with the U.S. Department of Energy, Office of Science, Office of High Energy Physics.

\appendix

\section{Dark Sector Thermalization Details}
\label{app:DS_thermalization_details}

A detailed calculation of the thermalization temperatures would require knowledge of the out-of-equilibrium distribution functions of the DCPs, $f_i(E)$. These are non-trivial to calculate in general~\cite{Binder:2021bmg,Du:2021jcj}, but simplifying assumptions can be made. We will assume that the distribution function follows equilibrium distribution of the SM, rescaled by the absolute abundance computed with Eq.~\eqref{eq:Y_boltz_FI},
\begin{align}
    f_X(E) \approx \left( \frac{n_X}{n^\text{eq}_X} \right) f^\text{eq}_X(E; T) \, .
    \label{eq:distribution_approximation}
\end{align}
This approximation captures two critical features. First, the number density calculated from integrating Eq.~\eqref{eq:distribution_approximation} is, by construction, the exact freeze-in number density. Second, the energy distribution of the DCPs is peaked near $T$, appropriate since the energy of the outgoing DCPs is set by the incoming charged SM particles.

This approximation is convenient because the relevant collision terms in the Boltzmann equations can be rescaled from their equilibrium values and written in terms of the usual thermally-averaged quantities. For example, the collision terms for $\phi \rightarrow \psi \chi$ and $\phi \phi \rightarrow A'A'$ are $R_{\phi \rightarrow \psi\chi} \approx n_\phi \left\langle \Gamma \right\rangle_{\phi \rightarrow \psi\chi}, R_{\phi \rightarrow A'} \approx n_\phi^2 \langle \sigma v \rangle_{\phi\phi \rightarrow A'A'}$, respectively, where $\langle \rangle$ represents a thermal average over equilibrium distributions at the SM temperature $T$. The thermal averages are computed in the usual way, e.g.,
\begin{align}
    & n^\text{eq}_\phi \, \langle \Gamma \rangle_{\phi \rightarrow \psi \chi} = \frac{m_\phi T}{4 \, ( 2\pi)^3} \sqrt{1 - \frac{(m_\psi - m_\chi)^2}{m_\phi^2}} \sqrt{1 - \frac{(m_\psi + m_\chi)^2}{m_\phi^2} } \mathscr{M}_{\phi \rightarrow \psi \chi}^2 \, K_1(m_\phi / T) \\
    & ( n^\text{eq}_\phi )^2 \, \langle \sigma v \rangle_{\phi \phi \rightarrow A'A'} = \frac{T}{16 \, (2 \pi)^5} \int_{s_\text{min}}^\infty \dd s \, \sqrt{s}  \sqrt{1 - \frac{4 M^2}{s}} \, \mathscr{M}^2_{\phi\phi \rightarrow A'A'} \, K_1(\sqrt{s} / T) \, .
\end{align}

The thermalization temperature of a given reaction set is then computed by equating the reaction rate to the Hubble rate for the slowest process. For example, for the first reaction set, $T_\text{th}^1$ can be computed by solving,
\begin{align}
    \text{min}\left\{ \langle\Gamma \rangle_{\phi \rightarrow \psi\chi}, n_\phi \, \langle \sigma v \rangle_{\phi\phi \rightarrow A'A'},  n_\psi \,\langle \sigma v \rangle_{\psi\psi \rightarrow A'A'}, n_\phi \, \langle \sigma v \rangle_{\phi\phi \rightarrow \chi\chi} \right\} \simeq H \, .
    \label{eq:thermalization_condition_1}
\end{align}
where the $n_X = s Y_X$ are computed from Eq.~\eqref{eq:Y_FI_final}. 

\section{Matrix Elements and Decay Rates}
\label{app:integrated_density_collision_terms}

The purpose of this appendix is to present all of the matrix elements and decay rates necessary to precisely compute the evolution of the DS energy density discussed in Sec.~\ref{subsec:DS_temperature_evolution}, and illustrated in Fig.~\ref{fig:xi_fig}. This involves including the contributions from the electroweak bosons, quarks and pions above and below the QCD scale, respectively, and also including plasmon decay~\cite{Dvorkin:2019zdi}.

We begin by discussing interactions with the electroweak bosons. Above the electroweak scale, the DS-SM interaction Lagrangian is given by~\cite{Bhattiprolu:2023akk},
\begin{align}
    \mathcal{L}_{\text{DS}-\text{SM}} = - \frac{\varepsilon}{2 \cos{\theta_W}} F'_{\mu \nu} B^{\mu \nu} \, ,
    \label{eq:app:L_int}
\end{align}
where $B_{\mu \nu} = \partial_\mu B_\nu - \partial_\nu B_\mu$ is the hypercharge gauge field strength tensor, $B_\mu$ is the hypercharge gauge field, and $\theta_W$ is the Weinberg angle, $\sin^2{\theta_W} \approx 0.23$~\cite{Tiesinga:2021myr}. Following Ref.~\cite{Fabbrichesi:2020wbt}, the field redefinition required to diagonalize the interaction in Eq.~\eqref{eq:app:L_int} is,

\begin{align}
    \begin{pmatrix} W^3_\mu \\ B_\mu \\ A'_\mu \end{pmatrix} \rightarrow \begin{pmatrix} \cos{\theta_W} & \sin{\theta_W} & -\sin{\theta_W} \varepsilon \\ -\sin{\theta_W} & \cos{\theta_W} & - \cos{\theta_W} \varepsilon \\ \tan{\theta_W} \varepsilon & 0 & 1 \end{pmatrix} \begin{pmatrix} Z_\mu \\ A_\mu \\ A'_\mu \end{pmatrix} \, ,
\end{align}
to $\mathcal{O}(\varepsilon)$. This field redefinition leads to the interactions~\cite{Bhattiprolu:2023akk},
\begin{align}
    \mathcal{L} \supset & -e \varepsilon A'_\mu J_\text{SM}^\mu - g' A'_\mu J_\text{DS}^{\prime\mu}  \nonumber \\
    &\quad- g' \varepsilon \tan{\theta_W} Z_\mu J^\mu_\text{DS} + i \varepsilon e \left[ F'^{\mu \nu}W^+_\mu W^-_\nu - F_+^{\mu \nu} A'_\mu W^-_\nu + F_-^{\mu \nu} A'_\mu W_+^\nu \right] \, ,
    \label{eq:app:L_int_2}
\end{align}
where $F_\pm^{\mu\nu} \equiv \partial^\mu W_{\pm}^\nu - \partial_\nu W_{\pm}^\mu$,  $J_\text{DS}^\mu \equiv J_\psi^{\prime\mu} + J_\phi^{\prime\mu}$ and $J_\psi^\mu, J_\phi^\mu$ are defined in the main text. The terms in the first line of Eq.~\eqref{eq:app:L_int_2} were discussed in Sec.~\ref{sec:model}, the first term in the second line appears due to the $A'_\mu \rightarrow A'_\mu + \tan\theta_W \varepsilon Z_\mu$ transformation, and the second term on the second line is from the cubic interaction in the $W$ boson kinetic term. 

The interaction Lagrangian in Eq.~\eqref{eq:app:L_int_2} is the generalization of Eq.~\eqref{eq:int_L} needed to compute all relevant matrix elements and decay rates, which we give below for different processes.

\vspace{1em}
\noindent\textbf{Lepton Annihilation}: Lepton annihilation to the DCPs, $\psi$ and $\phi$, proceeds via an $s$-channel diagram with $A'$ or $Z$ exchange. The spin-summed, angularly averaged matrix elements squared, Eq.~\eqref{eq:summed_averaged_M}, for these processes are given by,
\begin{align}
    & \mathscr{M}^2_{ff \rightarrow \psi\psi} = \frac{16}{3} e^4 \kappa^2 \left(1+\frac{2m_\psi^2}{s}\right) \Bigg[ q_f^2 \left(1+\frac{2m_f^2}{s} \right)  \nonumber \\ 
    & \hspace{2em}+\tan^2 \theta_W \frac{V_f^2s(s+m_f^2)+A_f^2s(s-4m_f^2)}{(s-m_Z^2)^2+m_Z^2 \Gamma_Z^2} -2 q_f V_f \tan \theta_W \frac{(s+2m_f^2)(s-m_Z^2)}{(s-m_Z^2)^2+m_Z^2 \Gamma_Z^2} \Bigg]
\end{align}
\begin{align}
    & \mathscr{M}^2_{ff \rightarrow \phi\phi} = \frac{4}{3} e^4 \kappa^2 \left(1-\frac{4m_\phi^2}{s}\right) \Bigg[ q_f^2 \left(1+\frac{2m_f^2}{s}\right)  \nonumber \\ 
    & \hspace{1.5em} +\tan^2 \theta_W \frac{V_f^2s(s+m_f^2)+A_f^2s(s-4m_f^2)}{(s-m_Z^2)^2+m_Z^2 \Gamma_Z^2} -2 q_f V_f \tan \theta_W \frac{(s+2m_f^2)(s-m_Z^2)}{(s-m_Z^2)^2+m_Z^2 \Gamma_Z^2} \Bigg] \, ,
\end{align}
where $2eV_{e,\mu,\tau} = -1 + 4 \sin^2\theta_W, \; 2eV_{\nu_e, \nu_\mu, \nu_\tau} = 1$ and $-2eA_{e,\mu,\tau} = 2eA_{\nu_e, \nu_\mu, \nu_\tau} = 1$ are the vector and axial couplings of the $Z$ to lepton pairs~\cite{Bhattiprolu:2023akk}, respectively, $m_Z \approx 91 \, \text{GeV}$ is the $Z$ mass, and $\Gamma_Z \approx 2 \, \text{GeV}$ is the $Z$ decay width.

\vspace{1em}
\noindent\textbf{Quark and Pion Annihilation}: In addition to lepton pair annihilation, at temperatures above the QCD scale, which we take to be $\Lambda_\text{QCD} \approx 200 \, \text{MeV}$, quarks can also annihilate via $s$-channel $A'$ and $Z$ exchange to DCPs with the matrix elements,
\begin{align}
    & \mathscr{M}^2_{ff \rightarrow \psi\psi} = 3 \times\frac{16}{3} e^4 \kappa^2 \left(1+\frac{2m_\psi^2}{s}\right) \Bigg[ q_f^2 \left(1+\frac{2m_f^2}{s}\right)  \nonumber \\ 
    & \hspace{2em} +\tan \theta_W^2 \frac{V_f^2s(s+m_f^2)+A_f^2s(s-4m_f^2)}{(s-m_Z^2)^2+m_Z^2 \Gamma_Z^2} -2 q_f V_f \tan \theta_W \frac{(s+2m_f^2)(s-m_Z^2)}{(s-m_Z^2)^2+m_Z^2 \Gamma_Z^2} \Bigg]
\end{align}
\begin{align}
    & \mathscr{M}^2_{ff \rightarrow \phi\phi}= 3 \times\frac{4}{3} e^4 \kappa^2 \left(1-\frac{4m_\phi^2}{s}\right) \Bigg[ q_f^2 \left(1+\frac{2m_f^2}{s}\right)  \nonumber \\ 
    & \hspace{2em} +\tan \theta_W^2 \frac{V_f^2s(s+m_f^2)+A_f^2s(s-4m_f^2)}{(s-m_Z^2)^2+m_Z^2 \Gamma_Z^2} -2 q_f V_f \tan \theta_W \frac{(s+2m_f^2)(s-m_Z^2)}{(s-m_Z^2)^2+m_Z^2 \Gamma_Z^2} \Bigg] \, ,
\end{align}
where $2eV_{u,c,t} = 1 - (8/3) \sin^2\theta_W, 2eV_{d,s,b} = -1 + (4/3) \sin^2\theta_W$ and $2eA_{u,c,t} = -2eA_{d,s,b} = 1$ are the vector and axial couplings of the $Z$ to quark pairs, respectively, and the overall factor of 3 is from the three quark flavors.

At temperatures below $\Lambda_\text{QCD}$ the quark annihilation processes do not proceed. Instead pions can annihilate to the DCPs. The dominant contribution is from charged pion annihilation through the $A'$, whose matrix elements are given by,
\begin{align}
    \mathscr{M}^2_{\pi^+ \pi^- \rightarrow \psi\psi} &= \frac{4}{3} e^4 \kappa^2 \left(1-\frac{4m_{\pi^\pm}^2}{s}\right) \left(1+\frac{2m_\psi^2}{s}\right)\\
\mathscr{M}^2_{\pi^+ \pi^- \rightarrow \phi\phi} &=\ \frac{1}{3} e^4 \kappa^2 \left(1-\frac{4m_{\pi^\pm}^2}{s}\right) \left(1-\frac{4m_\phi^2}{s}\right) \, ,
\end{align}
where $m_{\pi^{\pm}} \approx 140 \, \text{MeV}$.

\vspace{1em}
\noindent\textbf{$W$ Boson Annihilation}: $W$ bosons can also annihilate to the DCPs, with matrix elements given by~\cite{Bhattiprolu:2023akk},
\begin{align}
    \mathscr{M}^2_{W^+ W^- \rightarrow \psi\psi} &= \frac{1}{3} e^4 \kappa^2 \left(1-\frac{4m_{W}^2}{s}\right) \left(1+\frac{2m_\psi^2}{s}\right) \left(\frac{m_Z}{m_W}\right)^4 \frac{s^2+20s m_W^2+12 m_W^4}{(s-m_Z^2)^2+m_Z^2\Gamma_Z^2} \\
    \mathscr{M}^2_{W^+ W^- \rightarrow \phi\phi} &= \frac{1}{12} e^4 \kappa^2 \left(1-\frac{4m_{W}^2}{s}\right) \left(1-\frac{4m_\phi^2}{s}\right) \left(\frac{m_Z}{m_W}\right)^4 \frac{s^2+20s m_W^2+12 m_W^4}{(s-m_Z^2)^2+m_Z^2\Gamma_Z^2}
\end{align}
where $m_W \approx 80 \, \text{GeV}$ is the $W$ boson mass.

\vspace{1em}
\noindent\textbf{$Z$ Boson Decay}: Another type of contribution to the integrated energy density collision terms are decay processes. $R^\rho$ for the $Z$ decay processes are,
\begin{align}
    R^\rho_{Z \rightarrow\psi} & = \frac{3m_Z^3 T}{2\pi^2} K_2(m_Z/T) \left[ \frac{e^2 \kappa^2}{12 \pi} m_Z \tan^2{\theta_W} \sqrt{1-\frac{4m_\psi^2}{m_Z^2}} \left(1+\frac{2m_\psi^2}{m_Z^2} \right) \right] \\
    R^\rho_{Z \rightarrow\phi} & = \frac{3m_Z^3 T}{2\pi^2} K_2(m_Z/T) \left[ \frac{e^2 \kappa^2}{48 \pi} m_Z \tan^2{\theta_W} \left(1-4\frac{m_\phi^2}{m_Z^2} \right)^{3/2} \right] \, .
\end{align}
This contribution is important for $1~\textrm{GeV}\lesssim T \lesssim 100~\textrm{GeV}$ \cite{Chu:2011be}.

\vspace{1em}
\noindent\textbf{Plasmon Decay}: Lastly, for $T<1~\textrm{MeV}$, plasmon decay should be considered as well~\cite{Dvorkin:2019zdi,Chang:2019xva,Bhattiprolu:2023akk,Bhattiprolu:2024dmh}. $R^\rho$ for plasmon decay is given by,
\begin{equation}\label{eq:rhoplasmondecay}
    R^\rho_{\gamma^* \rightarrow \psi,\phi} = \frac{4 \pi}{(2 \pi)^3} \sum_\textrm{pol}\int_0^{k_\textrm{max}^\text{pol}} k^2 \dd k \,   \frac{\omega_\textrm{pol}}{e^{\omega_\text{pol}/T}-1}\, \Gamma^\textrm{pol}_{\gamma^* \rightarrow \psi\psi,\phi\phi} \,  \, ,
\end{equation}
where `pol' indicates the polarizations of the plasmon (2 transverse polarizations, $T$, and 1 longitudinal polarization $L$), $\omega_\text{pol} = \omega_\text{pol}(k)$ is the dispersion relation for the specific polarization, and $k_\text{max}^\text{pol}$ is the maximum maximum momentum for each polarization. The plasmon dispersion relations are found by solving,
\begin{eqnarray}
\omega_T^2&=&k^2 +\omega_p^2\frac{3\omega_T^2}{2\vst^2k^2}\left(1-\frac{\omega_T^2 -\vst^2k^2}{\omega_T^2}\frac{\omega_T}{2\vst k}\ln\left(\frac{\omega_T +\vst k}{\omega_T -\vst k}\right)\right) \, , \label{eq:dispersionT} \\
\omega_L^2&=&\op^2\frac{3\oml^2}{\vst^2 k^2}\left(\frac{\oml}{2\vst k}\ln\left(\frac{\oml +\vst k}{\oml -\vst k}\right)-1\right) \, .\label{eq:dispersionL}\,
\end{eqnarray}
where,
\begin{eqnarray}
\label{eq:omegaP}
\omega_p^2=\frac{2 e^2}{\pi^2}\int_0^\infty \dd p \, \frac{p^2}{E}\frac{(1 - v^2/3)}{e^{E/T}+1}~~~,~~~\omega_1^2=\frac{2 e^2}{\pi^2}\int_0^\infty \dd p \, \frac{p^2}{E} \frac{(5v^2/3 - v^4)}{e^{E/T}+1} \,, 
\end{eqnarray}
and $v=p/E$, $\vst=\omega_1/\op$, and $E=\sqrt{p^2+m_e^2}$. The plasmon decay rates are,
\begin{eqnarray}
    \Gamma^\textrm{pol}_{\gamma^* \rightarrow \psi\psi} &=& e^2 \kappa^2\frac{ Z_\textrm{pol}}{12 \pi}\sqrt{1-\frac{4m_\psi^2}{\omega_\textrm{pol}^2 - k^2}}\frac{(\omega_\text{pol}^2 - k^2)+2m_\psi^2}{\omega_\textrm{pol}} \, ,\\
    \Gamma^\textrm{pol}_{\gamma^* \rightarrow \phi\phi} &=& e^2 \kappa^2\frac{ Z_\textrm{pol}}{48 \pi}\sqrt{1-\frac{4m_\phi^2}{\omega_\textrm{pol}^2 - k^2}}\frac{(\omega_\text{pol}^2 - k^2)-4m_\phi^2}{\omega_\textrm{pol}} \, ,
\end{eqnarray}
with
\begin{align}
Z_T(k)&=\frac{2\omt^2 (\omt^2-\vst^2k^2)}{3\op^2\omt^2+(\omt^2+k^2)(\omt^2-\vst^2k^2)-2\omt^2(\omt^2-k^2)}\ , \\
\quad Z_L(k)&=\frac{2(\oml^2-\vst^2k^2)}{3\op^2-(\oml^2-\vst^2k^2)}\frac{\oml^2}{\oml^2-k^2}\,.
\end{align}
Finally, the maximum momentum for the longitudinal polarizations is,
\be
k_{\text{max}}^L&=&\op\left[\frac{3}{\vst^2}\lb \frac{1}{2\vst}\ln\lb\frac{1+\vst}{1-\vst}\rb-1\rb\right]^{1/2} \, ,
\ee
and the maximum momentum for the transverse polarizations is infinite.

\bibliographystyle{JHEP}
\bibliography{biblio.bib}

\end{document}